\documentclass[sigconf,authorversion]{acmart}

\usepackage{tikz}
\usepackage{amsmath,amsfonts,pifont}

\usepackage{booktabs}
\usepackage{array}
\newcolumntype{P}[1]{>{\centering\arraybackslash}p{#1}}
\usepackage{array}
\usepackage{tabularx, makecell}
\usepackage{multirow}
\newcolumntype{C}{>{\centering\arraybackslash}X}

\usepackage{pifont}

\usepackage{xspace}

\usepackage{enumitem}
\usepackage{caption}
\usepackage{subcaption}

\usepackage{xcolor}
\definecolor{LightGray}{gray}{0.9}

\usepackage{listings}
\definecolor{backcolour}{rgb}{0.95,0.95,0.92}
\lstdefinestyle{mystyle}{
    backgroundcolor=\color{backcolour},
    breakatwhitespace=true,         
    breaklines=true,
    numbers=left,
    tabsize=2
}
\acmConference[]{ }{ }{ }

\lstdefinelanguage{netunicorn}
{keywords={Pipeline, Client, pipeline, node, Nodes, filter, map, take, get, then, Experiment, deploy, execute, >>, |, ||, >+, +, &, push, if_, elif_, else, pop, for, in, range}, 
keywordstyle=\bfseries\ttfamily,
keywordstyle=[2]\ttfamily\bfseries,%
commentstyle=\ttfamily, %
stringstyle=\ttfamily, %
identifierstyle=\ttfamily,
emph={trafficAnomalyIPs}, 
emphstyle=\ttfamily\bfseries\color{red},
sensitive=true, alsoletter={0,1,2,3,4,5,6,7,8,9,-,>>,+,&,|,_},comment=[l][\footnotesize\sffamily\textbf]{\!}
}

\newcommand{\system}{netUnicorn\xspace}
\newcommand{\university}{\texttt{UCSB}\xspace}
\newcommand{\smartparagraph}[1]{\noindent{\bf #1}\ }
\newcommand{\rev}[1]{#1}

\setcopyright{none}
\renewcommand\footnotetextcopyrightpermission[1]{} %

\begin{document}

\date{}

\title[In Search of \system]{In Search of \system: A Data-Collection Platform to Develop Generalizable ML Models for Network Security Problems}
\subtitle{Extended version\\\href{https://netunicorn.cs.ucsb.edu}{\texttt{https://netunicorn.cs.ucsb.edu}}}

\author{Roman Beltiukov}
\email{rbeltiukov@ucsb.edu}
\orcid{0000-0001-8270-0219}
\affiliation{%
  \institution{UC Santa Barbara}
  \state{California}
  \country{USA}
}

\author{Wenbo Guo}
\email{henrygwb@purdue.edu}
\orcid{0000-0002-6890-4503}
\affiliation{%
  \institution{Purdue University}
  \state{Indiana}
  \country{USA}
}

\author{Arpit Gupta}
\email{agupta@ucsb.edu}
\orcid{0000-0002-6378-7440}
\affiliation{%
  \institution{UC Santa Barbara}
  \state{California}
  \country{USA}
}

\author{Walter Willinger}
\email{wwillinger@niksun.com}
\orcid{0000-0002-1384-8188}
\affiliation{%
  \institution{NIKSUN, Inc.}
  \state{New Jersey}
  \country{USA}
}

\renewcommand{\shortauthors}{Beltiukov, et al.}

\begin{sloppypar}
\begin{abstract}

The remarkable success of the use of machine learning-based solutions for network security problems has been impeded by the developed ML models' inability to maintain efficacy when used in different network environments exhibiting different network behaviors. This issue is commonly referred to as the generalizability problem of ML models. The community has recognized the critical role that training datasets play in this context and has developed various techniques to improve dataset curation to overcome this problem. Unfortunately, these methods are generally ill-suited or even counterproductive in the network security domain, where they often result in unrealistic or poor-quality datasets.

To address this issue, we propose a new closed-loop ML pipeline that leverages explainable ML tools to guide the network data collection in an iterative fashion. 
To ensure the data's realism and quality, we require that the new datasets should be endogenously collected in this iterative process, thus advocating for a gradual removal of data-related problems to improve model generalizability.
To realize this capability, we develop a data-collection platform, \system, that takes inspiration from the classic ``hourglass'' model and is implemented as its ``thin waist" to simplify data collection for different learning problems from diverse network environments.
The proposed system decouples data-collection intents from the deployment mechanisms and disaggregates these high-level intents into smaller reusable, self-contained tasks. 
We demonstrate how \system simplifies collecting data for different learning problems from multiple network environments and how the proposed iterative data collection improves a model's generalizability. 
\end{abstract}

\maketitle
\fancyhead{} 
\pagestyle{plain}

\section{Introduction}
\label{sec:intro}

Machine learning-based methods have outperformed existing rule-based approaches for addressing different network security problems, such as detecting DDoS attacks~\cite{kitsune}, malwares~\cite{invariant,atlas}, network intrusions~\cite{deeplog}, etc. 
However, their excellent performance typically relies on the assumption that the training and testing data are independent and identically distributed. Unfortunately, due to the highly diverse and adversarial nature of real-world network environments, this assumption does not hold for most network security problems. For instance, an intrusion detection model trained and tested with data from a specific environment cannot be expected to be effective when deployed in a different environment, where attack and even benign behaviors may differ significantly due to the nature of the environment. This inability of existing ML models to perform as expected in different deployment settings is known as \textit{generalizability problem}~\cite{Damour2020}, poses serious issues with respect to maintaining the models' effectiveness after deployment, and is a major reason why security practitioners are reluctant to deploy them in their production networks in the first place.

Recent studies (e.g.,~\cite{dos}) have shown that the quality of the training data plays a crucial role in determining the generalizability of ML models. In particular, in popular application domains of ML such as computer vision and natural language processing~\cite{van2001art,zhang2017mixup}, researchers have proposed several data augmentation and data collection techniques that are intended to improve the generalizability of trained models by enhancing the diversity and quality of training data~\cite{GUPTA2019466}. For example, in the context of image processing, these techniques include adding random noise, blurring, and linear interpolation. Other research efforts leverage open-sourced datasets collected by various third parties to improve the generalizability of text and image classifiers.

Unfortunately, these and similar existing efforts are not directly applicable to network security problems. For one, since the semantic constraints inherent in real-world network data are drastically different from those in text or image data, simply applying existing augmentation techniques that have been designed for text or image data is likely to result in unrealistic and semantically incoherent network data. Moreover, utilizing open-sourced data for the network security domain poses significant challenges, including the encrypted nature of increasing portions of the overall traffic and the fact that without detailed knowledge of the underlying network configuration, it is, in general, impossible to label additional data correctly. Finally, due to the high diversity in network environments and a myriad of different networking conditions, randomly using existing data or collecting additional data without understanding the inherent limitations of the available training data may even reduce data quality. As a result, there is an urgent need for novel data curation techniques that are specifically designed for the networking domain and aid the development of generalizable ML models for network security problems.

\rev{To address this need, we propose a new closed-loop ML pipeline (workflow) that focuses on training generalizable ML models for networking problems. Our proposed pipeline is a major departure from the widely-used standard ML pipeline~\cite{Damour2020} in two major ways. First, instead of obscuring the role that the training data plays in developing and evaluating ML models, the new pipeline elucidates the role of the training data. Second, instead of being indifferent to the black-box nature of the trained ML model, our proposed pipeline deliberately focuses on developing explainable ML models. To realize our new ML pipeline, we designed it using a closed-loop approach that leverages a novel data collection platform (called \system) in conjunction with state-of-the-art explainable AI (XAI) tools so as to be able to iteratively collect new training data for the purpose of enhancing the ability of the trained models to generalize. Here, during each iteration, the insights obtained from applying the employed explainability tools to the current version of the trained model are used to synthesize new policies for exactly what kind of new data to collect in the next iteration so as to combat generalizability issues affecting the current model.}

\rev{
In designing and implementing \system, the novel data collection platform that our proposed ML pipeline relies on, we leveraged state-of-the-art programmable data-plane targets, programmable network infrastructures, and different virtualization tools to enable flexible data collection at scale from disparate network environments and for different learning problems without network operators having to worry about the details of implementing their desired data collection efforts. 
This platform can be envisioned as representing the ``thin waist" of the classic hourglass model~\cite{10.1145/3274770}, where the different learning problems comprise the top layer and the different network environments constitute the bottom layer. 
To realize this ``thin waist" analog, \system supports a new programming abstraction that (i)~decouples the data-collection intents or policies (i.e., answering what data to collect and from where) from the mechanisms (i.e., answering how to collect the desired data on a given platform); and (ii)~disaggregates the high-level intents into self-contained and reusable subtasks.
}

\rev{In effect, our newly proposed ML pipeline advances the current state-of-the-art in ML model development by (1)~augmenting the standard ML pipeline with an explainability step that impacts how ML models are evaluated before being suggested for deployment, (2)~leveraging existing explainable AI (XAI) tools to identify issues with the utilized training data that may affect a trained model's ability to generalize, and (3)~using the insights gained from (2)~to inform the \system-enabled effort to iteratively collect new datasets for model training so as to gradually improve the generalizability of the models that are trained with these new datasets. 
A main difference between this novel closed-loop ML workflow and existing ``open-loop" ML pipelines is that the latter are either limited to using synthetic data for model training in their attempt to improve model generalizability or lack the means to collect data from network environments or for learning problems that differ from the ones that were specified for these pipelines in the first place. 
In this paper, we show that because of its ability to iteratively collect the ``right" training data from disparate network environments and for any given learning problem, our newly proposed ML pipeline paves the way for the development of generalizable ML models for networking problems.}

\smartparagraph{Contributions.}
This paper makes the following contributions:
\begin{itemize}
    \item {\bf An alternative ML pipeline.} 
    \rev{
    We propose a novel closed-loop ML pipeline that leverages a new data-collection platform in conjunction with state-of-the-art explainability (XAI) tools to enable iterative and informed data collection to gradually improve the quality of the data used for model training and thus boost the trained models' generalizability (\autoref{sec:background}). 
    }
    \item {\bf A new data-collection platform.} 
    \rev{We justify (\autoref{sec:justification}) and present the design and implementation (\autoref{ssec:thin-waist}) of \system, the new data-collection platform that is key to performing iterative and informed data collection for any given learning problem and from any network environment as part of our newly proposed closed-loop ML pipeline in practice.
    We made several design choices in \system to tackle the research challenges of realizing the ``thin waist'' abstraction.}

    \item 
    \rev{ \textbf{An extensive evaluation.}
    We demonstrate the capabilities of \system and the effectiveness of our newly proposed ML pipeline by (i)~considering various learning models for network security problems that have been studied in the existing literature and (ii)~evaluating them with respect to their ability to generalize (\autoref{sec:eval_iterative} and \autoref{sec:eval_platform}).
    }
    \item {\bf Artifacts.} We make the full source code of the system as well as the datasets used in this paper, publicly available (anonymously). Specifically, we have released three repositories: full source code of \system~\cite{p181system}, a repository of all discussed tasks and data-collection pipelines~\cite{p181library}, and other supplemental materials~\cite{p181suppl} (See \autoref{appendix:opensource}). 
\end{itemize}

We view the proposed ML pipeline and the new data-collection platform it relies on to be a promising first step toward developing  ML-based network security solutions that are generalizable and can, therefore, be expected to have a better chance of getting deployed in practice.
However, much work remains, and careful consideration has to be given to the network infrastructure used for data collection and the type of traffic observed in production settings before model generalizability can be guaranteed.

\section{Background and Problem Scope}
\label{sec:background}

\begin{figure*}[t]
    \centering
    \includegraphics[width=.9\textwidth]{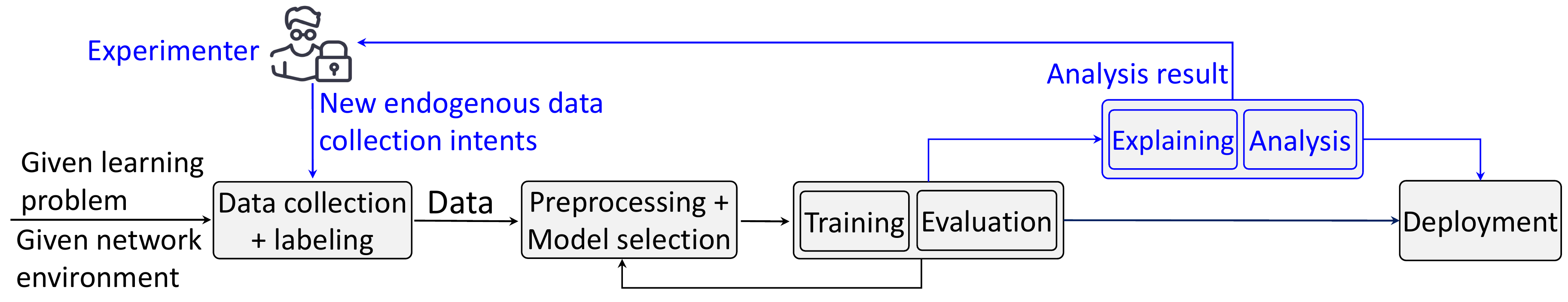}
    \caption{Overview of the existing (standard) and the newly-proposed (closed-loop) ML pipelines. 
    The components marked in blue are our proposed augmentations to the standard ML pipeline. 
    }
    \label{fig:overview_ml}
    \vspace{-10pt}
\end{figure*}

\subsection{Existing ML Pipeline for Network Security}

\smartparagraph{Key components.}
The standard ML pipeline (see \autoref{fig:overview_ml}) defines a workflow for developing ML artifacts and is widely used in many application domains, including network security.
To solve a learning problem (e.g., detecting DDoS attack traffic), the first step is to collect (or choose) labeled data, select a model design or architecture (e.g., random forest classifier), extract related features, and then perform model training using the training dataset. An independent and identically distributed (iid) evaluation procedure is then used to assess the resulting model by measuring its expected predictive performance on test data drawn from the training distribution.
The final step involves selecting the highest-performing model from a group of similarly trained models based on one or more performance metrics (e.g., F1-score). 
The selected model is then considered the ML-based solution for the task at hand and is recommended for deployment and being used or tested in production settings.

\smartparagraph{Data collection mechanisms.}
As in other application areas of ML, the collection of appropriate training data is of paramount importance for developing effective ML-based network security solutions.
In network security, the standard ML pipeline integrates two basic data collection mechanisms: \textit{real-world network data} collection and \textit{emulation-based network data} collection. 

In the case of real-world network data collection, data such as traffic-specific aspects are extracted directly (and usually passively) from a real-world target network environment. While this method can provide datasets that reflect pertinent attributes of the target environment, issues such as encrypted network traffic and user privacy considerations pose significant challenges to understanding the context and correctly labeling the data. Despite an increasing tendency towards traffic encryption 
~\cite{chatzoglou2022revisiting}, this approach still captures real-world networking conditions but often restricts the quality and diversity of the resulting datasets.

Regarding emulation-based network data collection, the approach involves using an existing or building one's own emulated environment of the target network and generating (usually actively) various types of attack and benign traffic in this environment to collect data. Since the data collector has full control over the environment, it is, in general, easy to obtain ground truth labels for the collected data. While created in an emulated environment, the resulting traffic is usually produced by existing real-world tools. Many widely used network datasets, including the still-used DARPA1998 dataset~\cite{darpa} and the more recent CIC-IDS intrusion detection datasets~\cite{cic} have been collected using this mechanism.

\subsection{Model Generalizability Issues}
\label{subsec:eval_problem}

Although existing emulation-based mechanisms have the benefit of providing datasets with correct labels, the training data is often riddled with problems that prevent trained models from generalizing, thus making them ill-suited for real-world deployment. 

There are three main reasons why these problems can arise. First, network data is inherently complex and heterogeneous, making it challenging to produce datasets that do not contain inductive biases. Second, emulated environments typically differ from the target environment -- without full knowledge of the target environment's configurations, it is difficult to accurately mimic it. The result is datasets that do not fully represent all the target environment's attributes. Third, shifting attack (or even benign) behavior is the norm, resulting in training datasets that become less representative of newly created testing data after the model is deployed.

These observations motivate considering the following concrete issues concerning the generalizability of ML-based network security solutions but note that there is no clear delineation between notions such as \textit{credible}, \textit{trustworthy} or \textit{robust} ML models and that the existing literature tends to blur the line between these (and other) notions and what we refer to as model generalizability.

\noindent
\underline{Shortcut learning.}
As discussed in~\cite{dos}, %
ML-based security solutions often suffer from shortcuts.
Here, shortcuts refer to encoded/inductive biases in a trained model that stem from false or non-causal associations in the training dataset~\cite{geirhos2020shortcut}. 
These biases can lead to a model not performing as desired in deployment scenarios, mainly because the test datasets from these scenarios are unlikely to contain the same false associations. 
Shortcuts are often attributable to data-collection issues, including how the data was collected (intent) or from where it was collected (environment).
Recent studies have shown that shortcut learning is a common problem for ML models trained with datasets collected from emulated networking environments.
For example,~\cite{trustee} found that the reported high F1-score for the VPN vs. non-VPN classification problem in~\cite{DraperGil2016CharacterizationOE} was due to a specific artifact of how this dataset was curated. 

\noindent
\underline{Out-of-distribution issues.}
Due to unavoidable differences between a real-world target environment and its emulated counterpart or subtle changes in attack and/or benign behaviors,
out-of-distribution (ood) data is another critical factor that limits model generalizability. 
The standard ML pipeline's evaluation procedure results in models that may appear to be well-performing, but their excellent performance can often be attributed to the models' innate ability for ``rote learning'', where the models cannot transfer learned knowledge to new situations.
To assess such models' ability to learn beyond iid data, purposefully curated ood datasets can be used. 

For network security problems, ood datasets of interest can represent different real-world network conditions (e.g., different user populations, protocols, applications, network technologies, architectures, or topologies) or different network situations (also referred to as distribution shift~\cite{quinonero2008dataset} or concept drift~\cite{Lu_2018}). 
For determining whether or not a trained model generalizes to different scenarios, it is important to select ood datasets that accurately reflect the different conditions that can prevail in those scenarios.

\subsection{Existing Approaches}
\label{sec:existing_approaches}
We can divide the existing approaches to improving a model's generalizability into two broad categories: (1)~Efforts for improving model selection, training, and testing algorithms; and (2)~Efforts for improving the training datasets. 
The first category focuses mainly on the later steps in the standard ML pipeline (see \autoref{fig:overview_ml}) that deal with the model's structure, the algorithm used for training, and the evaluation process. 
The second category is concerned with improving the quality of datasets used during model training and focuses on the early steps in the standard ML pipeline.

\smartparagraph{Improving model selection, training, and evaluation.}
The focal point of most existing efforts is either 
the model's structure (e.g., domain adaption~\cite{shankar2018generalizing,farahani2021brief} and multi-task learning~\cite{ruder2017overview,zhang2018overview}), or the training algorithm (e.g., few-shot learning~\cite{rivero2017grassmannian,goodfellow2016deep}), or the evaluation process (e.g., ood detection~\cite{jordaney2017transcend,yang2021cade}). 
However, they neglect the training dataset, 
mainly because it is in general assumed to be fixed and already given. 
While these efforts provide insights into improving model generalizability, studying the problem without the ability to actively and flexibly change the training dataset is difficult, especially when the given training dataset turns out to exhibit inductive biases, be noisy or of low quality, or simply be non-informative for the problem at hand~\cite{GUPTA2019466}. See \autoref{sec:related} for a more detailed discussion about existing model-based efforts and how they differ from our proposed approach described below.

\smartparagraph{Improving the training dataset.}
Data augmentation is a passive method for synthesizing new or modifying existing training datasets and is widely used in the ML community to improve models' generalizability.
Technically, data augmentation methods leverage different operations (e.g., adding random noise~\cite{van2001art}, using linear interpolations~\cite{zhang2017mixup} or more complex techniques) to synthesize new training samples for different types of data such as images~\cite{van2001art,shorten2019survey}, text~\cite{zhang2017mixup}, or tabular data~\cite{smote,smotestudy}. 
However, using such passive data-generation methods for the network security domain is inappropriate or counterproductive because they often result in unrealistic or even semantically meaningless datasets~\cite{gepperth2020survey}.
For example, since network protocols usually adhere to agreed-upon standards, they constrain various network data in ways that such data-generation methods cannot ensure without specifically incorporating domain knowledge.
Furthermore, various network environments can induce significant differences in observed communication patterns, even when using the same tools or considering the same scenarios~\cite{DHOOGE2020102564},  by influencing data characteristics (e.g., packet interarrival times, packet sizes, or header information) and introducing unique network conditions or patterns.

\vspace{-5pt}
\subsection{Limitations of Existing Approaches}

From a network security domain perspective, these existing approaches miss out on two aspects that are intimately related to improving a model's ability to generalize: (1)~Leveraging insights from model explainability tools, and (2)~ensuring the realism of collected training datasets.

\smartparagraph{Using explainable ML techniques.}
To better scrutinize an ML model's weaknesses and understand model errors, we argue that an additional explainability step that relies on recent advances in explainable ML should be added to the standard ML pipeline to improve the ML workflow for network security problems~\cite{https://doi.org/10.48550/arxiv.1911.01058,PETCH2022204, trustee,guo2018lemna}. 
The idea behind adding such a step is that it enables taking the output of the standard ML pipeline, extracting and examining a carefully-constructed white-box model in the form of a decision tree, and then scrutinizing it for signs of blind spots in the output of the standard ML pipeline. 
If such blind spots are found, the decision tree and an associated summary report can be consulted to trace their root causes to aspects of the training dataset and/or model specification that led the output to encode inductive biases.

\smartparagraph{Ensuring realism in collected training datasets.}
To beneficially study model generalizability from the training dataset perspective, we posit that for the network security domain, the collection of training datasets should be done \textit{endogenously} or \textit{in vivo}; that is, performed or taking place within the network environment of interest. 
Given that network-related datasets are typically the result of intricate interactions between different protocols and their various embedded closed control loops, accurately reflecting these complexities associated with particular deployment settings or traffic conditions requires collecting the datasets from within the network.

\subsection{Our Approach in a Nutshell}
We take a first step towards a more systematic treatment of the model generalizability problem and propose an approach that (1)~uses a new closed-loop ML pipeline and (2)~calls for running this pipeline in its entirety multiple times, each time with a possibly different model specification but always with \textit{a different training dataset} compared to the original one. 
Here, we use a newly-proposed closed-loop ML pipeline (\autoref{fig:overview_ml}) that differs from the standard pipeline by including an explanation step.
Also, each new training dataset used as part of a new run of the closed-loop ML pipeline is assumed to be endogenously collected and not exogenously manipulated.

\rev{
The collection of each new training dataset is informed by a root cause analysis of identified inductive bias(es) in the trained model. 
This analysis leverages existing explainability tools that researchers have at their disposal as part of the closed-loop pipeline’s explainability step. In effect, such an informed data-collection effort promises to enhance the quality of the given training datasets by gradually reducing the presence of inductive biases that are identified by our approach, thus resulting in trained models that are more likely to generalize. Note, however, that our proposed approach \textit{does not guarantee} model generalizability. Instead, by eliminating identified inductive biases in the form of shortcuts and ood data, our approach enhances a model's generalizability capabilities. Also, note that our focus in this paper is not on designing novel model explainability methods but rather on applying available techniques from the existing literature. In fact, while we are agnostic about which explainability tools to use for this step, we recommend the application of global explainability tools such as Trustee~\cite{trustee} over local explainability techniques (e.g., ~\cite{guo2018lemna,shap,lime,Vasi_2022, xnids}), mainly because the former are in general more powerful and informative with respect to faithfully detecting and identifying root causes of inductive biases compared to the latter. However, as shown in~\autoref{sec:eval_iterative} below, either of these two types of methods can shed light on the nature of a trained model’s inductive biases.}

Our proposed approach differs from existing approaches in several ways. 
First, it reduces the burden on the user or domain expert to select the ``right'' training dataset apriori. Second, it calls for the collection of training datasets that are endogenously generated and where explainability tools guide the decision-making about what ``better" data to collect. Third, it proposes using multiple training datasets, collected iteratively (in a fail-fast manner), to combat the underspecification of the trained models and thus enhance model generalizability. 
In particular, it recognizes that an ``ideal'' training dataset may not be readily available in the beginning and argues strongly against attaining it through exogenous means.

\begin{figure}[t]
     \centering
         \includegraphics[width=\linewidth]{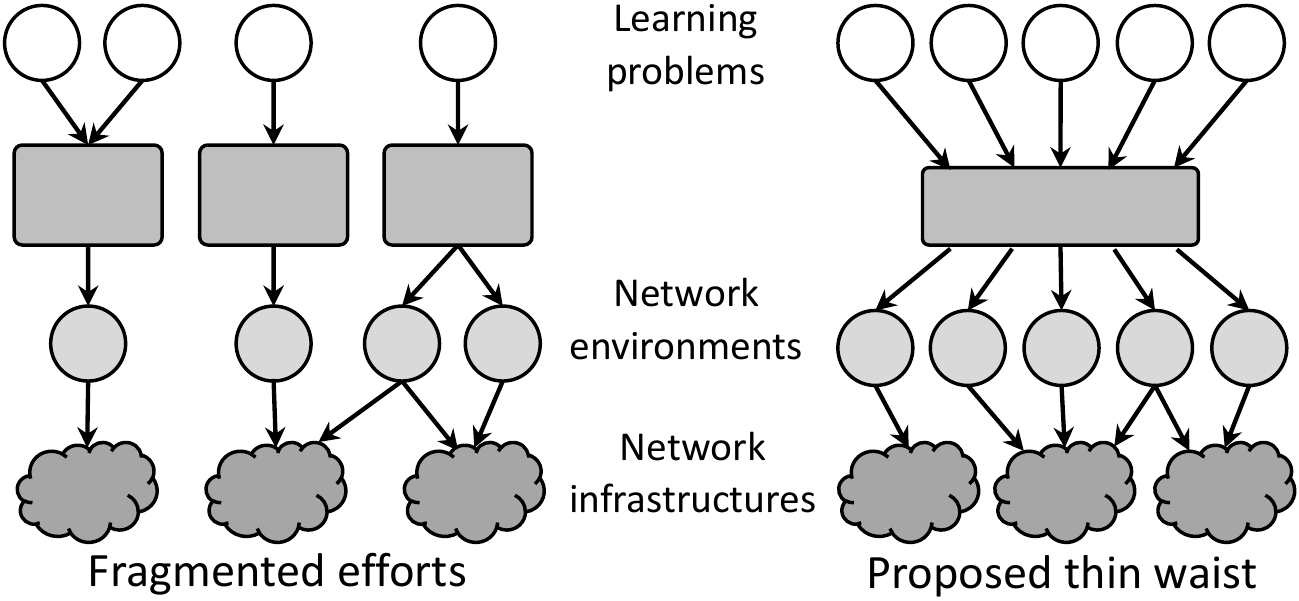}
        \caption{\system vs. existing data collection efforts.}
        \label{fig:ours_previous_dt}
        \vspace{-15pt}
\end{figure}

\section{On ``in vivo'' Data-Collection}
\label{sec:justification}

In this section, we discuss some of the main issues with existing data-collection efforts and describe our proposed approach to overcome their shortcomings.

\subsection{Existing Approaches}
\smartparagraph{Data collection operations.}
We refer to collecting data for a learning problem from a specific network environment (or domain) as a data-collection \textit{experiment}.
We divide such a data-collection experiment into three distinct operations.
(1)~\textit{Specification:} expressing the intents that specify what data to collect or generate for the experiment. (2)~\textit{Deployment:} bootstrapping the experiment by translating the high-level intents into target-specific commands and configurations across the physical or virtual data-collection infrastructure and implementing them. (3)~\textit{Execution:} orchestrating the experiment to collect the specified data while handling different runtime events (e.g., node failure, connectivity issues, etc.). 
Here, the first operation is concerned with ``what to collect," and the latter operations deal with ``how to collect" this data.  

\smartparagraph{The ``fragmentation'' issue.}
Existing data-collection efforts are inherently \textit{fragmented}, i.e., they only work for a specific learning problem and network environment, emulated using one or more network infrastructures (\autoref{fig:ours_previous_dt}).
Extending them to collect data for a new learning problem or from a new network environment is challenging.
For example, consider the data-collection effort for the video fingerprinting problem~\cite{beautyburst}, where the goal is to fingerprint different videos for video streaming applications (e.g., YouTube) using a stream of encrypted network packets as input. 
Here, the data-collection intent is to start a video streaming session and collect the related packet traces from multiple end hosts that comprise a specific target environment.
The deployment operation entails developing scripts that automate setting up the computing environment (e.g., installing the required selenium package) at the different end hosts. 
The execution operation requires developing a runtime system to start/stop the experiments and handle runtime events such as node failure, connectivity issues, etc. 

\smartparagraph{Lack of modularity.}
In addition to being one-off in nature, existing approaches to collecting data for a given learning problem are also monolithic. 
That is, being highly problem-specific, there is, in general, no clear separation between experiment specification and mechanisms.
An experimenter must write scripts that realize the data-collection intents (e.g., start/stop video streaming sessions, collect pcaps, etc.), deploy these scripts to one or more network infrastructures, and execute them to collect the required data. 
Given this monolithic structure, existing data collection approaches~\cite{beautyburst} cannot easily be extended so that they can be used for a different learning problem, such as inferring QoE~\cite{netmicroscope,10.1145/3523230.3523232, requet} or for a different network environment, such as congested environments (e.g., hotspots in a campus network) or high-latency networks (e.g., networks that use GEO satellites as access link). 

\smartparagraph{\rev{Disparity between virtual and physical infrastructures.}}
\rev{
While a number of different network emulators and simulators are currently available to researchers~\cite{mininet,ns3,mahimahi,pantheon}, it is, in general, difficult or impossible to write experiments that can be seamlessly transferred from a virtual to a physical infrastructure and back. This capability is particularly appealing in view of the fact that virtual infrastructures provide the ability to quickly iterate on data collection and test various network conditions, including conditions that are complex in nature and, in general, difficult to achieve in physical infrastructures. Due to the lack of this capability, experimenters often end up writing experiments for only one of these infrastructures, creating different (typically simplified) experiment versions for physical test beds, or completely rewriting the experiments to account for real-world conditions and problems (e.g., node and link failures, network synchronization)
}

\smartparagraph{Missed opportunity.}
Together, these observations highlight a missed opportunity for researchers who now have access to different network infrastructures. 
The list includes NSF-supported research infrastructures, such as EdgeNet~\cite{edgenet}, ChiEdge~\cite{chiedge}, Fabric~\cite{fabric}, PAWR~\cite{pawr}, etc., as well as on-demand infrastructure offered by different cloud services providers, such as AWS~\cite{aws}, Azure~\cite{azure}, Digital Ocean~\cite{digitalocean}, GCP~\cite{gcp}, etc.
This rich set of network infrastructures can aid in emulating diverse and representative network environments for data collection.
\vspace{-3pt}

\subsection{An ``Hourglass'' Design to the Rescue}
The observed fragmented, one-off, and monolithic nature of how training datasets for network security-related ML problems are currently collected motivates a new and more principled approach that aims at lowering the threshold for researchers wanting to collect high-quality network data.
Here, we say a training dataset is of high quality if the model trained using this dataset is not obviously prone to inductive biases and, therefore, likely to generalize. 

\smartparagraph{Our hourglass model.} 
Our proposed approach takes inspiration from the classic ``hourglass'' model~\cite{10.1145/3274770}, a layered systems architecture that, in our case, consists of designing and implementing a ``thin waist" that enables collecting data for different learning problems (hourglass' top layer) from a diverse set of possible network environments (hourglass' bottom layer). 
In effect, we want to design the thin waist of our hourglass model in such a way that it accomplishes three goals:
(1)~allows us to collect a specified training dataset for a given learning problem from network environments emulated using one or more supported network infrastructures, 
(2)~ensures that we can collect a specified training set for each of the considered learning problems for a given network environment, and 
(3)~facilitates experiment reproducibility and shareability.

\smartparagraph{Requirements for a ``thin waist''.}
Realizing this hourglass model's thin waste requires developing a flexible and modular data-collection platform that supports two main functionalities:
(1)~\textit{decoupling} data-collection intents (i.e., expressing \textbf{what} to collect and from \textbf{where}) from mechanisms (i.e., {\bf how} to realize these intents); and (2)~\textit{disaggregating} intents into independent and reusable tasks.

The required first functionality allows the experimenter to focus on the experiment's intent without worrying about how to implement it. As a result, expressing a data-collection experiment does not require re-doing tasks related to deployment and execution in different network environments. 
For instance, to ensure that the learning model for video fingerprinting is not overfitted to a specific network environment, collecting data from different environments, such as congested campus networks or cable- and satellite-based home networks, is important. 
Not requiring the experimenter to specify the implementation details simplifies this process.

Providing support for the second functionality allows the experimenter to reuse common data-collection intents and mechanisms for different learning problems. 
For instance, while the goal for QoE inference and video fingerprinting may differ, both require starting and stopping video streaming sessions on an end host. 

Ensuring these two required functionalities makes it easier for an experimenter to iteratively improve the data collection intent, addressing apparent or suspected inductive biases that a model may have encoded and may affect the model's ability to generalize.

\section{Realizing the ``Thin Waist'' Idea}
\label{ssec:thin-waist}
To achieve the desired ``thin waist'' of the proposed hourglass model, we develop a new data-collection platform, \system. 
We identify two distinct stakeholders for this platform: (1)~\textit{experimenters} who express data-collection intents, and (2)~\textit{developers} who develop different modules to realize these intents. 
In~\autoref{ssec:abstractions}, we describe the programming abstractions that \system considers to satisfy the ``thin'' waist requirements, and
in ~\autoref{ssec:dni}, we show how \system realizes these abstractions while ensuring fidelity, scalability, and extensibility. 

\subsection{Programming Abstractions}
\label{ssec:abstractions}
To satisfy the second requirement (\textit{disaggregation}), \system allows experimenters to disaggregate their intents into distinct pipelines and tasks.  
Specifically, \system offers experimenters {\tt Task} and {\tt Pipeline} abstractions.  
Experimenters can structure data collection experiments by utilizing multiple independent pipelines.
Each pipeline can be divided into several processing stages, where each stage conducts self-contained and reusable tasks. 
In each stage, the experimenter can specify one or more tasks that \system will execute concurrently. 
Tasks in the next stage will only be executed once all tasks in the previous stage have been completed.

To satisfy the first requirement, \system offers a unified interface for all tasks. 
To this end, it relies on abstractions that concern specifics of the computing environment (e.g., containers, shell access, etc.) and executing target (e.g., ARM-based Raspberry Pis, AMD64-based computers, OpenWRT routers, etc.) and allows for flexible and universal task implementation.

To further decouple intents from mechanisms, \system's API exposes the {\tt Nodes} object to the experimenters. 
This object abstracts the underlying physical or virtual infrastructure as a pool of data-collection nodes. 
Here, each node can have different static and dynamic attributes, such as type (e.g., Linux host, PISA switch), location (e.g., room, building), resources (e.g., memory, storage, CPU), etc. 
An experimenter can use the {\tt filter} operator to select a subset of nodes based on their attributes for data collection.
Each node can support one or more compute environments, where each environment can be a shell (command-line interpreter), a Linux container (e.g., Docker~\cite{docker}), a virtual machine, etc.
\system allows users to map pipelines to these nodes using the {\tt Experiment} object and {\tt map} operator.
Then, experimenters can deploy and execute their experiments using the {\tt Client} object. 
\autoref{tab-api} in the appendix summarizes the key components of \system's API.

\smartparagraph{Illustrative example.} 
\label{sec:patator}
To illustrate with an example how an experimenter can use \system's API to express the data-collection experiment for a learning problem, we consider the bruteforce attack detection problem.
For this problem, we need to realize three pipelines, where the different pipelines perform the key tasks of running an HTTPS server, sending attacks to the server, and sending benign traffic to the server, respectively.
The first pipeline also needs to collect packet traces from the HTTPS server. 

\autoref{listing:patator} shows how we express this experiment using \system. 
Lines 1-6 show how we select a host to represent a target server, start the HTTPS server, perform PCAP capture, and notify all other hosts that the server is ready.
Lines 8-16 show how we can take hosts from different environments that will wait for the target server to be ready and then launch a bruteforce attack on this node.
Lines 18-26 show how we select hosts that represent benign users of the HTTPS server.
Finally, lines 28-32 show how we combine pipelines and hosts into a single experiment, deploy it to all participating infrastructure nodes, and start execution.

Note that in \autoref{listing:patator} we omitted task definitions and instantiation, package imports, client authorization, and other details to simplify the exposition of the system.

\begin{figure}[t]
\begin{lstlisting}[language=netUnicorn,basicstyle=\footnotesize, numbers=left,xleftmargin=2em,frame=single,label=listing:patator,
framexleftmargin=2.0em, captionpos=b, caption=Data collection experiment example for the HTTPS bruteforce attack detection problem. We have omitted task instantiations and imports to simplify the exposition. 
]
# Target server
h1 = Nodes.filter('location', 'azure').take(1)
p1 = Pipeline()
     .then(start_http_server)
     .then(start_pcap)
     .then(set_readiness_flag)

# Malicious hosts
h2 = [
  Nodes.filter('location', 'campus').take(40),
  Nodes.filter('location', 'aws').take(40),
  Nodes.filter('location', 'digitalocean').take(40),
]
p2 = Pipeline()
     .then(wait_for_readiness_flag)
     .then(patator_attack)

# Benign hosts
h3 = [
  Nodes.filter('location', 'campus').take(40),
  Nodes.filter('location', 'aws').take(40),
  Nodes.filter('location', 'digitalocean').take(40),
]
p3 = Pipeline()
     .then(wait_for_readiness_flag)
     .then(benign_traffic)

e = Experiment()
    .map(p1, h1)
    .map(p2, h2)
    .map(p3, h3)
Client().deploy(e).execute(e)
\end{lstlisting}
\vspace*{-15pt}
\end{figure}

\subsection{System Design}
\label{ssec:dni}

The \system compiles high-level intents, expressed using the proposed programming abstraction, into target-specific programs. 
It then deploys and executes these programs on different data-collection nodes to complete an experiment.
\system is designed to realize the high-level intents with \textit{fidelity}, minimize the inherent computing and communication overheads ({\em scalability}), and simplify supporting new data-collection tasks and infrastructures for developers (\textit{extensibility}).

\smartparagraph{Ensuring high fidelity.}
\system is responsible for compiling a high-level experiment into a sequence of target-specific programs.
We divide these programs into two broad categories for each task: deployment and execution.
The deployment definitions help configure the computing environment to enable the successful execution of a task. 
For example, executing the {\tt YouTubeWatcher} task requires installing a Chromium browser and related extensions.
Since successful execution of each specified task is critical for satisfying the fidelity requirement, \system must ensure that the computing environment at the nodes is set up for a task before execution.

\smartparagraph{Addressing the scalability issues.}
To execute a given pipeline, a system can control deployment and execution either at the task- or pipeline-level granularity. 
The first option entails the deployment and execution of the \textit{task} and then reporting results back to the system before executing the next \textit{task}. It ensures fidelity at the task granularity and allows the execution of pipelines even with tasks with contradicting requirements (e.g., different library versions). However, since such an approach requires communication with core system services, it slows the completion time and incurs additional computing and network communication overheads.

Our system implements the second option, running all the setup programs before marking a \textit{pipeline} ready for execution and then offloading the task flow control to a node-based executor that reports results only at the end of the pipeline. It allows for optimization of environment preparation (e.g., configure a single docker image for distribution) and time overhead between tasks, and also reduces network communication while offering only ``best-effort'' fidelity for pipelines.

\smartparagraph{Enabling extensibility.}
Enabling extensibility calls for simplifying how a developer can add a new task, update an existing task for a new target, or add a new physical or virtual infrastructure. 
Note that the \system's extensibility requirement targets developers and not experimenters.

\noindent
\underline{Simplify adding and updating tasks.}
An experimenter specifies a task to be executed in a pipeline. 
The \system chooses a specific implementation of this task.
This may require customizing the computing environment, which can vary depending on the target (e.g., container vs shell of OpenWRT router). 
For example, a Chromium browser and specific software must be installed to start a video streaming session on a remote host without a display. 
The commands to do so may differ for different targets. 
The system provides a base class that includes all necessary methods for a task.

Developers can extend this base class by providing their custom subclasses with the target-specific {\tt run} method to specify how to execute the task for different types of targets. 
This allows for easy extensibility because creating a new task subclass is all that is needed to adapt the task to a new computing environment.

\noindent
\underline{Simplify adding new infrastructures.}
To deploy data-collection pipelines, send commands, and send/receive different events and data to/from multiple nodes in the underlying infrastructure, \system requires an underlying deployment system. 

One option is to bind \system to one of the existing deployment (orchestration) systems, such as Kubernetes~\cite{k8s}, SaltStack~\cite{saltproject}, Ansible~\cite{ansible}, 
or others for all infrastructures. 
However, requiring a physical infrastructure to support a specific deployment system is disruptive in practice.
Network operators managing a physical infrastructure are often not amenable to changing their deployment system as it would affect other supported services. 

Another option is to support multiple deployment systems. 
However, we need to ensure that supporting a new deployment system does not require a major refactoring of \system's existing modules. 
To this end, \system introduces a separate connectivity module that abstracts away all the connectivity issues from the \system's other modules (e.g., runtime), offering seamless connectivity to infrastructures using multiple deployment systems. 
Each time developers want to add a new infrastructure that uses an unsupported deployment system, they only need to update the connectivity manager --- simplifying extensibility.

\begin{figure}[t]
    \centering
    \includegraphics[width=0.9\linewidth]{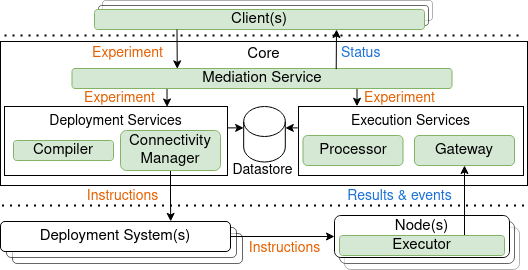}
    \caption{Architecture of the proposed system.  \textmd{Green-shaded boxes show all the implemented services.}}
    \label{fig:architecture}
    \vspace{-15pt}
\end{figure}

\subsection{Prototype Implementation}
Our implementation of \system is shown in \autoref{fig:architecture}.
Our implementation embraces a service-oriented architecture~\cite{richards2015software} and has three key components: \textit{client(s)}, \textit{core}, and \textit{executor(s)}. 
Experimenters use local instances of \system's \textit{client} to express their data-collection experiments. 
Then, \system's \textit{core} is responsible for all the operations related to the compilation, deployment, and execution of an experiment. 
For each experiment, \system's core deploys a target-specific \textit{executor} on all related data-collection nodes for running and reporting the status of all the programs provided by \system's core. 

The \system's core offer three main service groups: mediation, deployment, and execution services. 
Upon receiving an experiment specification from the client, the {\tt mediation service} requests the {\tt compiler} to extract the set of setup configurations for each distinct (pipeline, node-type) pair, which it uploads to the local PostgreSQL database. 
After compilation, the {\tt mediation service} requests the connectivity manager to ship this configuration to the appropriate data-collection nodes and verify the computing environment. 
In the case of docker-based infrastructures, this step is performed locally, and the configured docker image is uploaded to a local docker repository. 
The {\tt connectivity-manager} uses an infrastructure-specific deployment system (e.g., SaltStack~\cite{saltproject}) to communicate with the data-collection nodes.

After deploying all the required instructions, the {\tt mediation service} requests the connectivity manager to instantiate a target-specific {\tt executor} for all data-collection nodes. 
The {\tt executor} uses the instructions shipped in the previous stage to execute a data-collection pipeline. 
It reports the status and results to \system's {\tt gateway} and then adds them to the related table in the SQL database via the {\tt processor}. 
The {\tt mediation service} retrieves the status information from the database to provide status updates to the experimenter(s). 
Finally, at the end of an experiment, the {\tt mediation service} sends cleanup scripts (via {\tt connectivity-manager}) to each node---ensuring the reusability of the data-collection infrastructure across different experiments.

\section{Evaluation: Closed-loop ML Pipeline}
\label{sec:eval_iterative}

\begin{table*}[!ht]
\centering
\caption{{Number of LLoC changes, data points, and F1 scores across different environments and iterations.}
}
\label{tab:eval_shortcuts}
\resizebox{.8\textwidth}{!}{%
\begin{tabular}{c|c|c|c|c|c|c}
& \multicolumn{2}{c|}{Iteration \#0 (initial setup)} & \multicolumn{2}{c|}{Iteration~1 } & \multicolumn{2}{c}{Iteration~2} 
\\
\hline
LLoCs & \multicolumn{2}{c|}{\textbf{80}} & \multicolumn{2}{c|}{\textbf{+10}} & \multicolumn{2}{c}{\textbf{+20}} 
\\
\hline
& \university-0 (train) & {\tt multi-cloud} (test)
& \university-1 (train) & {\tt multi-cloud} (test)
& \university-2 (train) & {\tt multi-cloud} (test)
\\                           
\hline
MLP & 1.0 & 0.56 & 0.97 (-0.03) & 0.62 (+0.06) & 0.88 (-0.09) & \textbf{0.94 (+0.38)} 
\\
GB  & 1.0 & 0.61 & 1.0 (+0.00) & 0.61 (+0.00) & 0.92 (-0.08) & \textbf{0.92 (+0.31)} 
\\
RF  & 1.0 & 0.58 & 1.0 (+0.00) & 0.69 (+0.11) & 0.97 (-0.03) & \textbf{0.93 (+0.35)} \\
\rev{TN}  & \rev{1.0} & \rev{0.66} & \rev{0.97 (-0.03)} & \rev{0.78 (+0.12)} & \rev{0.92 (-0.05)} & \textbf{\rev{0.95 (+0.29)}} 
\end{tabular}%
}
\end{table*}

\begin{figure*}[!ht]
     \centering
     \begin{subfigure}[b]{0.3\linewidth}
         \centering
         \includegraphics[width=\textwidth]{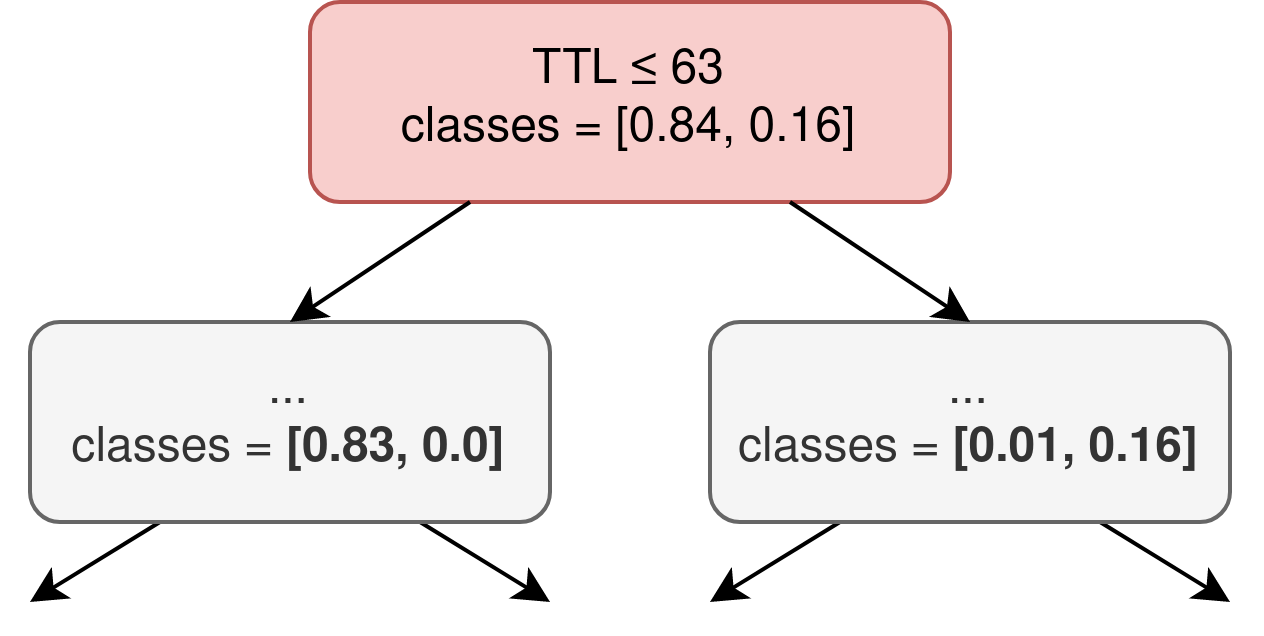}
         \caption{Iteration \#0: top branch is a shortcut.}
         \label{subfig:tree_iterative_1}
     \end{subfigure}
     \hfill
     \begin{subfigure}[b]{0.3\linewidth}
         \centering
         \includegraphics[width=\textwidth]{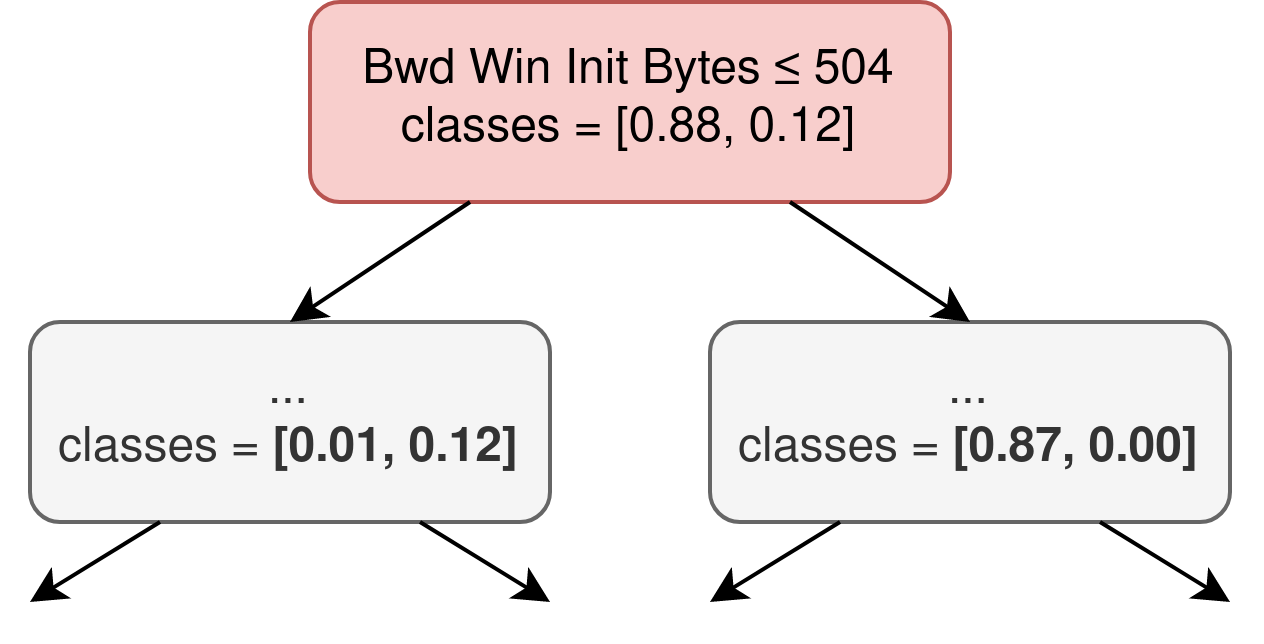}
         \caption{Iteration \#1: top branch is a shortcut.}
         \label{subfig:tree_iterative_2}
     \end{subfigure}    
     \hfill
     \begin{subfigure}[b]{0.3\linewidth}
         \centering
         \includegraphics[width=\textwidth]{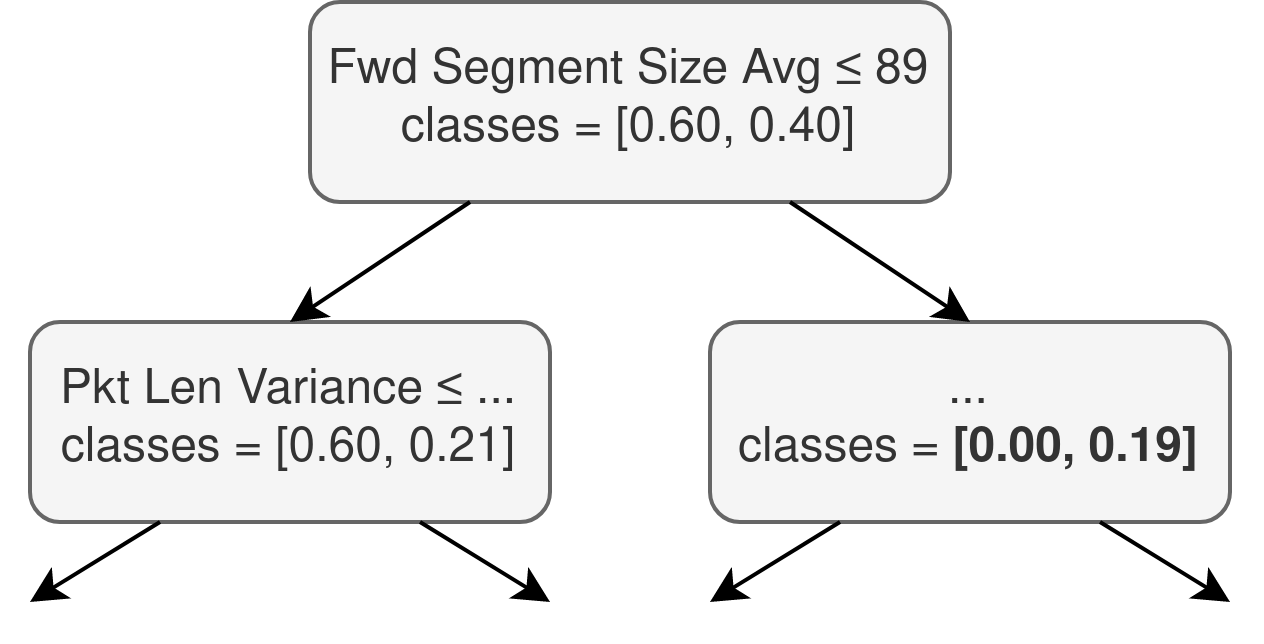}
         \caption{Iteration \#2: no obvious shortcut.}
         \label{subfig:tree_iterative_3}
     \end{subfigure}
     \hfill
        \caption{Decision trees generated using Trustee~\cite{trustee} across the three iterations. We highlight the nodes that are indicators for shortcuts in the trained model.}%
        \label{fig:tree_iterative}
\vspace{-10pt}
\end{figure*}

In this section, we demonstrate how our proposed closed-loop ML pipeline helps to improve model generalizability. 
Specifically, we seek to answer the following questions: 
\ding{182}~Does the proposed pipeline help in identifying and removing shortcuts?
\ding{183}~How do models trained using the proposed pipeline perform compared to models trained with existing exogenous data augmentation methods?
\ding{184}~Does the proposed pipeline help with combating ood issues?

\subsection{Experimental Setup}
\label{ssec:experimental_setup}

To illustrate our approach and answer these questions, we consider the bruteforce example mentioned in \autoref{ssec:abstractions} and first describe the different choices we made with respect to the ML pipeline and the iterative data-collection methodology.

\smartparagraph{Network environments.}
We consider three distinct network environments for data collection: a \university network, a hybrid \university{\tt-cloud} setting, and a {\tt multi-cloud} environment.

The \university network environment is emulated using a programmable data-collection infrastructure PINOT~\cite{pinot}. This infrastructure is deployed at a campus network and consists of multiple (40+) single-board computers (such as Raspberry Pis) connected to the Internet via wired and/or wireless access links. These computers are strategically located in different areas across the campus, including the library, dormitories, and cafeteria. In this setup, all three types of nodes (i.e., target server, benign hosts, and malicious hosts) are selected from end hosts on the campus network.
The \university{\tt-cloud} environment is a hybrid network that combines programmable end hosts at the campus network with one of three cloud service providers: AWS, Azure, or Digital Ocean.\footnote{Unless specified otherwise, we host the target server on Azure for this environment.} In this setup, we deploy the target server in the cloud while running the benign and malicious hosts on the campus network.
Lastly, the {\tt multi-cloud} environment is emulated using all three cloud service providers with multiple regions. We deploy the target server on Azure and the benign and malicious hosts on all three cloud service providers.

\smartparagraph{Data collection experiment.}
The data-collection experiment involves three pipelines, namely target, benign, and malicious. Each of these pipelines is assigned to different sets of nodes depending on the considered network environment. 
The target pipeline is responsible for deploying a public HTTPS endpoint with a real-world API that requires authentication for access. 
Additionally, this pipeline utilizes \textit{tcpdump} to capture all incoming and outgoing network traffic. 
The benign pipeline emulates valid usage of the API with correct credentials, while the malicious pipeline attempts to obtain the service's data by brute-forcing the API using the Patator~\cite{patator} tool and a predefined list of commonly used credentials~\cite{seclists}.

\smartparagraph{Data pre-processing and feature engineering.}
We used CICFlowMeter~\cite{cicflowmeter} to transform raw packets into a feature vector of 84 dimensions for each unique connection (flow). These features represent flow-level summary statistics (e.g., average packet length, inter-arrival time, etc.) and are widely used in the network security community~\cite{cicids2017, zhou2019evaluation, DraperGil2016CharacterizationOE, cuzzocrea2017tor}.

\smartparagraph{Learning models.}
\rev{
We train four different learning models.
Two of them are traditional ML models, i.e., Gradient Boosting (GB)~\cite{natekin2013gradient}, Random Forest (RF)~\cite{breiman2001random}.
The other two are deep learning-based methods: 
Multi-layer Perceptron (MLP)~\cite{goodfellow2016deep}, and attention-based TabNet model (TN)~\cite{arik2020tabnet}. 
These models are commonly used for handling tabular data such as CICFlowMeter features~\cite{tabular22,grinsztajn2022treebased}. 
}

\smartparagraph{Explainability tools.} \rev{To examine a model trained with a given training dataset for the possible presence of inductive biases such as shortcuts or ood issues, our newly proposed ML pipeline requires an explainability step that consists of applying existing model explainability techniques, be they global or local in nature, but what technique to use is left to the discretion of the user.}

\rev{We illustrate this step by first applying a global explainability method. In particular, our method-of-choice is the recently developed tool Trustee~\cite{trustee}, but other global model explainability techniques could be used as well, including PDP plots~\cite{10.1214/aos/1013203451}, ALE plots~\cite{apley2019visualizing}, and others~\cite{nori2019interpretml,molnar2020interpretable}. Our reasoning for using the Trustee tool is that for any trained black-box model, it extracts a high-fidelity and low-complexity decision tree that provides a detailed explanation of the trained model's decision-making process. Together with a summary report that the tool provides, this decision tree is an ideal means for scrutinizing the given trained model for possible problems such as shortcuts or ood issues.}

\rev{To compare, we also apply local explainability tools to perform the explainability step. More specifically, we consider the two well-known techniques, LIME~\cite{lime} and SHAP~\cite{shap}. These methods are designed to explain a model's decision for individual input samples and thus require analyzing the explanations of multiple inputs to make conclusions about the presence or absence of model blind spots such as shortcuts or ood issues. While users are free to replace LIME or SHAP with more recently developed tools such as xNIDS~\cite{xnids} or their own preferred methods, they have to be mindful of the efforts each method requires to draw sound conclusions about certain non-local properties of a given trained model (e.g., shortcut learning).}

\subsection{Identifying and Removing Shortcuts}
\label{subsec:eval_iterative_shortcut}

To answer \ding{182}, we consider a setup where a researcher curates training datasets from the \university environment and aims at developing a model that generalizes to the {\tt multi-cloud} environment (i.e., unseen domain). 

\smartparagraph{Initial setup (iteration \#0).}
We denote the training data generated from this experiment as \university{\tt-0}. 
\autoref{tab:eval_shortcuts} shows that while all three models have a perfect training performance, they all have low testing performance (errors are mainly false positives). \rev{We first used our global explanation method-of-choice, Trustee, to extract the decision tree of the trained models.
As shown in \autoref{fig:tree_iterative}, the top node is labeled with the separation rule ($TTL \le 63$) and the balance between the benign and malicious samples in the data (``classes’’). Subsequent nodes only show the class balance after the split.
}

\rev{
From \autoref{subfig:tree_iterative_1}, we conclude that all four models use almost exclusively the TTL (time-to-live) feature to discriminate between benign and malicious flows, which is an obvious shortcut. Note that the top parts of Trustee-extracted decision trees were identical for all four models.
When applying the local explanation tools LIME and SHAP to explain 100 randomly selected input samples, we found that these explanations identified TTL as the most important feature in all 100 samples. While consistent with our Trustee-derived conclusion, these LIME- or SHAP-based observations are necessary but not sufficient to conclusively decide whether or not the trained models learned a TTL-based shortcut strategy and further efforts would be required to make that decision.}

To understand the root cause of this shortcut, we checked the \university infrastructure and noticed that almost all nodes used for benign traffic generation have the exact same TTL value due to a flat structure of the \university network.
This observation also explains why most errors are false positives, i.e., the model treats a flow as malicious if it has a different TTL from the benign flows in the training set. 
Existing domain knowledge suggests that this behavior is unlikely to materialize in more realistic settings such as the {\tt multi-cloud} environment. 
Consequently, we observe that models trained using the \university{\tt-0} dataset perform poorly on the unseen domain; i.e., they generalize poorly.

\smartparagraph{Removing shortcuts (iteration \#1).}
To fix this issue, we modified the data-collection experiment to use a more diverse mix of nodes for generating benign and malicious traffic and collected a new dataset, \university{\tt-1}.
However, this change only marginally improved the testing performance for all three models (\autoref{tab:eval_shortcuts}).
Inspection of the corresponding decision trees shows that all the models use the ``Bwd Init Win Bytes'' feature for discrimination, which appears to be yet another shortcut. Again, we observed that all trees generated by Trustee from different black-box models have identical top nodes.
\rev{Similar, our local explanation results obtained by LIME and SHAP also point to this feature as being the most important one across the analyzed samples.}

More precisely, this feature quantifies the TCP window size for the first packet in the backward direction, i.e., from the attacked server to the client. 
It acts as a flow control and reacts to whether the receiver (i.e., HTTP endpoint) is overloaded with incoming data. 
Although it could be one indicator of whether the endpoint is being brute-force attacked, it should only be weakly correlated with whether a flow is malicious or benign.
Given this reasoning and the poor generalizability of the models, we consider the use of this feature to be a shortcut.

\smartparagraph{Removing shortcuts (iteration \#2).}
To remove this newly identified shortcut, we refined the data-collection experiment.
First, we created a new task that changes the workflow for the Patator tool. 
This new version uses a separate TCP connection for each bruteforce attempt and has the effect of slowing down the bruteforce process.
Second, we increased the number of flows for benign traffic and the diversity of benign tasks. Using these changes, we collected a new dataset, \university{\tt-2}.

\autoref{tab:eval_shortcuts} shows that the change in data-collection policy significantly improved the testing performance for all models. 
We no longer observe any obvious shortcuts in the corresponding decision tree. 
Moreover, domain knowledge suggests that the top three features (i.e., ``Fwd Segment Size Average'', ``Packet Length Variance'', and ``Fwd Packet Length Std'') are meaningful and their use can be expected to accurately differentiate benign traffic from repetitive brute force requests.
\rev{Applying the local explanation methods LIME and SHAP also did not provide any indications of obvious additional shortcuts.}
Note that although the models appear to be shortcut-free, we cannot guarantee that the models trained with these diligently curated datasets do not suffer from other possible encoded inductive biases. Further improvements of these curated datasets might be possible but will require more careful scrutiny of the obtained decision trees and possibly more iterations.

\begin{table}[t]
    \centering
    \caption{F1 score of models trained using our approach (i.e., leveraging \system) vs. models trained with datasets collected from the \university network by exogenous methods (i.e., without using \system).}
    \resizebox{.99\linewidth}{!}{
        \begin{tabular}{c|cccc|cccc|cccc}
        & \multicolumn{4}{c|}{Iteration \#0} & \multicolumn{4}{c|}{Iteration \#1} & \multicolumn{4}{c}{\textbf{Iteration \#2}} \\ 
        & MLP & GB & RF & \rev{TN} & MLP & GB & RF & \rev{TN} & MLP & GB & RF & \rev{TN} \\ 
        \hline
        \rev{Naive Aug.} & \rev{0.51} & \rev{0.57} & \rev{0.56} & \rev{0.53} & \rev{0.73} & \rev{0.67} & \rev{0.71} & \rev{0.82} & \rev{-} & \rev{-} & \rev{-} & \rev{-} \\
        Noise Aug.  & 0.66 & 0.68 & 0.67 & \rev{0.66} & 0.72 & 0.83 & 0.76 & \rev{0.82} & - & - & - & \rev{-} \\ 
        Feature Drop & 0.74 & 0.55 & 0.72 & \rev{\textbf{0.87}} & \textbf{0.91} & 0.58 & 0.63 & \rev{\textbf{0.89}} & - & - & - & \rev{-} \\ 
        SYMPROD & 0.66 & 0.71 & 0.67 & \rev{0.41} & 0.69 & 0.66 & 0.75 & \rev{0.67} & \textbf{0.94} & \textbf{0.93} &\textbf{ 0.95} & \rev{\textbf{0.96}} \\ 
        \hline
        \textbf{Our approach} & & & & & & & & &
        \textbf{0.94} & \textbf{0.92} & \textbf{0.95} & \rev{\textbf{0.95}} \\ 
        \end{tabular}
    }
    \label{table:compare}
\end{table}

\subsection{Comparison with Exogeneous Methods}
\label{subsec:eval_iterative_compare}
To answer \ding{183}, we compare the performance of the model trained using \university{\tt-2} (i.e., the dataset curated after two rounds of iterations) with that of models trained with datasets modified by means of existing exogenous methods. Specifically, we consider the following methods: 
\begin{enumerate}
    \item \textbf{\rev{Naive augmentation.}} \rev{We use a naive data collection strategy that does not apply the extra explanation step that our newly proposed ML pipeline includes to identify training data-related issues. The strategy simply collects more data using the initial data-collection policy. It is an ablation study demonstrating the benefits of including the explanation step in our new pipeline. Here, for each successive iteration, we double the size of the training dataset.} 
    \item \textbf{Noise augmentation.} This popular data augmentation technique consists of adding suitable chosen random uniform noise~\cite{maharana2022review} to the identified skewed features in each iteration.
    \rev{ 
    Here, for iteration \#0, we use integer-valued uniformly-distributed random samples from the interval $[-1; +1]$ for TTL noise augmentation, and for iteration \#1, we similarly use integer-valued uniformly-distributed samples from the interval $[-5; +5]$ for noise augmentation of the feature ``Bwd Init Win Bytes".}

    \item \textbf{Feature drop.} This method simply drops a specified skewed feature from the dataset in each iteration. 
    \rev{In our case, we drop the identified skewed feature for all training samples in each training dataset.}

    \item \textbf{SYMPROD.} SMOTE~\cite{smote} is a popular augmentation method for tabular data that applies interpolation techniques to synthesize data points to balance the data across different classes.
    Here we utilize a recently considered version of this method called SYMPROD~\cite{symprod} and 
    \rev{augment each training set by adding the number of rows necessary for restoring class balance ($proportion=1$).}

\end{enumerate}

\noindent

We apply these methods to the three training datasets curated from the campus network in the previous experiment. 
For \university{\tt-0} and \university{\tt-1}, we use the two identified skewed features for adding noise or dropping features altogether. 

\rev{Note that since we did not identify any skewed features in the last iteration, we did not apply any noise augmentation and feature drop techniques in this iteration and did not collect more data for the naive data augmentation method. 
}

As shown in \autoref{table:compare}, the models trained using these exogenous methods perform poorly in all iterations when compared to our approach.
This highlights the main benefit we gain from applying our proposed closed-loop ML pipeline for iterative data collection and model training. 
\rev{  
In particular, it demonstrates that the explanation step in our proposed pipeline adds value. While doing nothing (i.e., naive data augmentation) is clearly not a worthwhile strategy, applying either noise augmentation or SYMPROD can potentially compromise the semantic integrity of the training data, making them ill-suited for addressing model generalizability issues for network security problems.
}

\subsection{Combating ood-specific Issues}
\label{subsec:eval_iterative_OOD}

\begin{table}[t]
    \caption{The testing F1 score of the models before and after retraining with malicious traffic generated by Hydra.}
    \resizebox{.7\linewidth}{!}{%
    \centering
    \begin{tabular}{c|cccc|c}
        \multirow{2}{*}{} & \multirow{2}{*}{MLP}  & \multirow{2}{*}{GB} & \multirow{2}{*}{RF} & \multirow{2}{*}{\rev{TN}} & \multirow{2}{*}{Avg} \\
         & & & & & \\ \hline
        \multicolumn{1}{c|}{Before retraining} & \multicolumn{1}{c}{0.87} & 0.81 & 0.86 & \rev{0.83} & 0.84 \\ 
        \multicolumn{1}{c|}{After retraining} & \multicolumn{1}{c}{\textbf{0.93}} & \textbf{ 0.96} & \textbf{0.91} & \rev{\textbf{0.91}} & \textbf{0.93} \\ 
    \end{tabular}
    }
    \label{table:ood}
\end{table}

\begin{table}[t]
    \centering
        \caption{The F1 score of models trained using only \university data or data from \university and {\textbf{\university}-{\bf cloud}} infrastructures.}
    \resizebox{.75\linewidth}{!}{%
    \begin{tabular}{c|cc|cc}
        & \multicolumn{2}{c|}{\textbf{\university}} & \multicolumn{2}{c}{\textbf{\university}-{\bf cloud}} \\
        & Training & Test & Training & Test
        \\                            
        \hline
        MLP & 0.88 & 0.94 & \textbf{0.95 (+0.07)} & \textbf{0.95 (+0.01)} \\
        GB  & 0.92 & 0.92 & \textbf{0.96 (+0.04)} & \textbf{0.95 (+0.03)} \\
        RF  & 0.97 & 0.93 & \textbf{0.96 (-0.01)} & \textbf{0.97 (+0.04)} \\
        \rev{TN} & \rev{0.83} & \rev{0.95} & \rev{\textbf{0.84 (+0.01)}} & \rev{\textbf{0.96 (+0.01)}} 
    \end{tabular}%
    }
    \label{tab:multi_infra}
    \vspace{-10pt}
\end{table}

To answer \ding{184}, we consider two different scenarios: \textit{attack adaptation} and \textit{environment adaptation}.

\smartparagraph{Attack adaptation.} We consider a setup where an attacker changes the tool used for the bruteforce attack, i.e., uses Hydra~\cite{hydra} instead of Patator. 
To this end, we use \system to generate a new testing dataset from the \university infrastructure with Hydra as the bruteforce attack. 
\autoref{table:ood} shows that the model's testing performance drops significantly (to 0.85 on average). 
We observe that this drop is because of the model's reduced ability to identify malicious flows, which indicates that changing the attack generation tool introduces oods, although they belong to the same attack type.

To address this problem, we modified the data generation experiment to collect attack traffic from both Hydra and Patator in equal proportions.
This change in the data-collection experiment only required \textbf{6} LLoC.
We retrain the models on this dataset and observe significant improvements in the model's performance on the same test dataset after retraining (see \autoref{table:ood}). 

Note that we only test one type of oods where the evolved attack still has the same goal and functionality.
However, an attack can also evolve into another attack with a different goal, resulting in ood samples with new labels.
Here, we leverage ensemble models and human analysis to identify the ood case. While it may be possible to identify ood issues using more automated methods that are motivated by findings obtained from applying global explainability tools, 
we plan to revisit this problem in our future work.

\begin{figure}[t]
     \centering
     \begin{subfigure}[b]{0.49\linewidth}
         \centering
         \includegraphics[width=\textwidth]{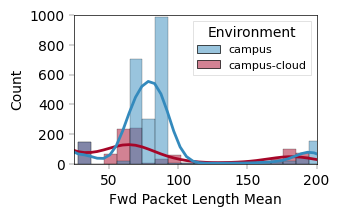}
     \end{subfigure}
     \hfill
     \begin{subfigure}[b]{0.49\linewidth}
         \centering
         \includegraphics[width=\textwidth]{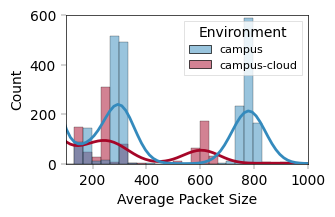}
     \end{subfigure}
     \hfill
        \caption{Distributions of several features across two different environments: \university and {\textbf{\university}-{\bf cloud}}}
        \label{fig:feature_dist}
\vspace{-15pt}
\end{figure}

\smartparagraph{Environment adaptation.}
We consider testing the model we developed in the \university environment in the unseen {\tt multi-cloud} environment as a different instance of an ood issue that is due to possible feature distribution differences. To address this issue, we use the \university{\tt-cloud} environment for data collection. As expected, we observe differences in the distributions for some of the features across the two environments (see \autoref{fig:feature_dist}). 
\autoref{tab:multi_infra} shows the performance of the models trained using only the data from the \university environment compared to the ones that use data from both the \university and \university{\tt-cloud} environments. Notably, as \university{\tt-cloud} is more similar to the {\tt multi-cloud} environment than the \university environment, the models trained with the \university{\tt-cloud} data show improvements in their performance under the test settings.

\section{Evaluation: \system}
\label{sec:eval_platform}
We answer if \system lowers the threshold for data collection for:
\ding{185}~different learning problems for a given network environment?
\ding{186}~a given learning problem from different environments, emulated using one or more network infrastructures? and
\ding{187}~iteratively calibrating the data collection intents for a given learning problem and environment? 
We also demonstrate \ding{188}~how well does \system scale for larger data-collection infrastructures, especially the ones equipped with relatively low-end devices, such as RPis?

\subsection{Experimental Setup}
\smartparagraph{Learning problems.}
Besides the \textit{HTTP bruteforce attack detection} problem, we explore two more learning problems for this experiment, namely \textit{video fingerprinting} and \textit{advanced persistent threats detection} (APTs). In the case of the first additional example, the learning problem is to fingerprint videos for web-based streaming services, such as YouTube, that adopt variable bitrates~\cite{beautyburst}.
Previous work~\cite{beautyburst} did not evaluate the proposed learning model under realistic network conditions. 
Thus, to collect meaningful data for this problem, we use a network of end hosts in the \university infrastructure to collect a training dataset for five different YouTube videos.\footnote{Each video is identified with a unique URL.} Specifically, our data-collection intent is specified by the following sequence of tasks: start packet capture, watch a YouTube video in headless mode for 30 seconds, and stop packet capture. We repeat this sequence ten times for each video in a shuffled order and combine it into a single pipeline, where at the end, we upload the collected data to our server.

Regarding the second additional example, the learning problem, in this case, is to identify the hosts that some APTs have compromised. 
To generate data for this learning problem, we write an experiment that mimics the behavior of a compromised host. Specifically, our data-collection intent is as follows: find active hosts using {\tt Ping}, check if port 443 is opened for active hosts (identified in the previous stage) with {\tt PortScan}, and then for each host with open 443 port launch four different attacks in parallel: {\tt CVE20140160} (Heartbleed), {\tt CVE202141773} (Apache 2.4.49 Path), {\tt CVE202144228} (Log4J), and {\tt Patator} (HTTP admin endpoint bruteforce using the Patator tool). The ML pipeline creates a ``semi-realistic'' training dataset by combining actively generated attack traffic with passively collected packet traces from a border router of a production network, such as the \university network.\footnote{Note, in theory, we could use \system to actively collect the benign traffic for this learning problem in addition to the attack traffic. However, generating representative benign traffic for a large and complex enterprise network will require a more complex data-collection infrastructure than the one we use for evaluation. \autoref{sec:discussion} discusses this issue in greater detail.}
We then use this dataset for model training.
Note, here we assume that we know the attacker's playbook; that is, the goal, in this case, is not to demonstrate a realistic attack playbook but to demonstrate that \system simplifies generating attack traffic for a given APT attack playbook.

\smartparagraph{Network environments.}
\system enables emulating network environments for data collection using one or more physical/virtual infrastructures. 
Previously, we used a SaltStack-based infrastructure at \university and multiple clouds to emulate various network environments: \university, \university{\tt-cloud}, and {\tt multi-cloud}.
In this experiment, we implement a connector to another infrastructure, Azure Container Instances (ACI) to expand cloud-based environments with serverless Docker containers. During the experiments, containers were dynamically created in multiple regions and used for pipeline execution.
Overall, \system currently supports six different deployment system connectors (see~\autoref{table:deploymentsystems} in \autoref{appendix:connectors}).

\smartparagraph{Baseline.}
To the best of our knowledge, none of the existing platforms/systems offer the desired extensibility, scalability, and fidelity for data collection (see~\autoref{sec:related} for more details).  
To illustrate how \system simplifies data collection efforts, we consider baselines that directly configure three different deployment/orchestration systems. Specifically, we consider the following deployment systems as baselines: {\bf Kubernetes}, \textbf{SaltStack}, and {\bf Azure Container Instances (ACI)}. For each data-collection experiment, we explicitly compose different tasks to realize different data-collection pipelines, create pipeline-specific docker images, and use existing tools (e.g., \textit{kubectl}) to map and deploy these pipelines to different nodes.

\subsection{Simplifying Data Collection Effort}
We now demonstrate how \system simplifies data collection for:

\smartparagraph{Different learning problems for a given network environment (\ding{185}).}
\autoref{tab:multiproblems} reports the effort in expressing the data-collection experiments for the three learning problems for the \university network. We observe that \system only requires 17-35 LLoCs to express the data-collection intent. The \university network infrastructure uses SaltStack as the deployment system, and we observe that it takes 113-237~LLoC (around 5-13~$\times$ more effort) to express and realize the same data-collection intents without \system. 

The key enabler here is the set of self-contained tasks that realize different data-collection activities. For each learning problem, \autoref{tab:multiproblems} quantifies the overhead of specifying new tasks unique to the problem at hand. Even taking the overheads of expressing these tasks into consideration, collecting the same data from \university network without \system requires around 2-3~$\times$ more effort. 

Overall, we implemented around twenty different tasks to bootstrap \system (see \autoref{table:appendixtaskloc} (in \autoref{appendix:tasks}) for more details). 
The total development effort for the bootstrapping was around \textbf{900 LLoCs }.
Though this bootstrapping effort is not insignificant, we posit that this effort amortizes over time as this repository of reusable and self-contained tasks will facilitate expressing increasingly disparate data-collection experiments.

\begin{table}[t]
\centering
\caption{LLoCs to implement different problems using \system and other deployment systems. Here, the three learning problems are (1)~Bruteforce detection, (2)~video fingerprinting, and (3)~APT detection.}
\resizebox{\linewidth}{!}{
\begin{tabular}{c|c|ccc}
\multirow{2}{5em}{\textbf{Learning\\Problems}} & \multicolumn{1}{c|}{\textbf{\system}} & \multicolumn{3}{c}{\textbf{Other Deployment Systems}}\\
& Experiment (Tasks) & Kubernetes & SaltStack & ACI \\ \hline
1 & \textbf{21}~(18) & 74 
& 113 
& 61 \\
2 & \textbf{35}~(115) & 161 & 237 & 179 \\
3 & \textbf{17}~(120) & 151 & 232 & 176 \\
\hline
\multicolumn{2}{l|}{LLoC Ratio for Experiments + Tasks} & $1-2\times$ & $2-3\times$ & $1-2\times$ \\
\multicolumn{2}{l|}{LLoC Ratio for Experiments} & $3-9\times$ & $5-13\times$ & $3-10\times$ \\
\end{tabular}
}
\label{tab:multiproblems}
\vspace{-15pt}
\end{table}

\smartparagraph{Given learning problem from multiple network environments (\ding{186}).}
As we discussed before, \system is inherently extensible, i.e., it can use different sets of network infrastructures to emulate disparate network environments for data collection. With \system, changing an existing data-collection experiment to collect data from a new set of network infrastructure(s) requires changing only a few LLoCs (2-3 for the examples in~\autoref{tab:multiproblems}). In contrast, collecting the data for the HTTP Bruteforce detection problem from a cloud infrastructure (ACI) and a Kubernetes cluster requires writing additional 61 and 74 LLoCs, respectively. This effort is even more intense for video fingerprinting and APT detection problems. 

The key enabler for simplifying data collection across one or more network infrastructures is \system's extensible {\tt connectivity-manager} that can interface with multiple deployment systems via a system of connectors. 
In \autoref{table:deploymentsystems}, we enumerated all the implemented connectors and corresponding logical lines of code (LLoC) for each implementation. 
Note that this bootstrapping is a one-time effort, and these connectors can be reused across multiple physical infrastructures that are managed using either of the supported deployment systems (e.g., SaltStack, Kubernetes, etc.). 

\smartparagraph{Iterative data collection (\ding{187}).}
To iteratively modify data collection intents, the system should allow flexibility in both pipeline modifications and environment changes. We implemented the experiment, described in \autoref{sec:eval_iterative}, using \system, for all three environments (\university, \university{\tt-cloud}, and {\tt multicloud}). We report the combined LLoCs for experiment definitions and tasks implementations in \autoref{tab:eval_shortcuts}. As we reused previously implemented connectors, we do not report their LLoC in the table.

The table shows that the overhead for iterative updates is minimal. While this overhead may also be minimal for more conventional (platform- and problem-specific) solutions, \system's abstractions allow for seamless integration of many other platforms, thus providing a means to increase the diversity of the collected datasets further and, in turn, a model's generalizability capabilities.

\subsection{Scaling Data Collection}
\label{subsec:scaling}
To quantify the computing and memory overheads of \system's core and executors (\ding{188}), we measure the wall time or elapsed time as a proxy for CPU cycles and use a Python-based memory profiler~\cite{memory-profiler}, respectively. 
Our results show that the executor running on a low-end node such as a Raspberry Pi incurs a computing overhead of approximately \textbf{1 second per stage} and \textbf{0.13 seconds per task} while consuming less than \textbf{21 MB of memory}. 
Meanwhile, \system's core incurs a computing overhead of around five seconds for deployment and 20 seconds for execution in a 20-node infrastructure while consuming less than \textbf{417 MB of memory}. The details of these experiments can be found in \autoref{appendix:measurements}. 

 \section{Discussion}
\label{sec:discussion}

\smartparagraph{More learning problems.}
\rev{While not implemented in this paper, we envision that the \system platform can be used for a wide range of different network security problems, such as network censorship~\cite{279954,281442,257178}, website fingerprinting~\cite{281438,277132}, Tor traffic analysis~\cite{10.1145/3548606.3560604}, and others. Many of these problems involve an active measurement component for data collection, labeling, or communication and would benefit from \system-provided capabilities such as (i) running experiments that require the simultaneous use of different infrastructures and (ii) facilitating the reproducibility and shareability of experiments. To demonstrate this benefit, we used \system to implement a multi-vantage point validation of the Let’s Encrypt ACME challenge~\cite{272232} and refer the reader to \autoref{appendix:letsencrypt} for further details. We provide additional evidence for the practicability and versatility of \system and its use as part of our newly-proposed ML pipeline by describing in \autoref{additional_iterative} the application of our approach to two additional real-world security problems, namely Heartbleed detection and OS fingerprinting.}

\smartparagraph{Usability and Realism.}
First, a critical step in our proposed method is that we require domain experts to articulate data collection intents. 
As demonstrated in \autoref{sec:eval_iterative}, it is often possible to generate appropriate intents with the help of explainable ML models. 
Our platform design further simplifies the process of translating intents into action, ensuring the usability of our proposed method.
Second, our data collection follows an emulation-based mechanism that enables accurate labeling. 
With our proposed iterative approach, we can eliminate biases from the collected data. 
Additionally, our platform significantly lowers the threshold for gathering data from multiple environments, enhancing the diversity of the data collected.
As demonstrated in \autoref{sec:eval_iterative}, the data we collected is realistic and representative and can improve the generalizability of trained models in various environments.

\smartparagraph{Limitations of the proposed approach.}

\noindent\underline{Active data collection.} 
Our approach uses endogenously generated (labeled) network data from actual network environments. 
We note that it may also be possible to improve a model's generalizability by means of carefully selected and exogenously generated (passive) data from a production network, but such an approach is beyond the scope of this paper. 

\noindent \underline{Feature pre-processing.}
Curating training datasets entails both data collection and pre-processing.
Since data pre-processing remains the same for different versions of the collected data that result from our iterative approach, it poses no problems for the desired ``thin waist'' of \system's design. 
In this paper, we utilized the CICFlowmeter for pre-processing, which worked well for all considered learning problems. While we readily acknowledge that there is more to data pre-processing than CICFlowmeter, we leave the exploration of alternative pre-processing (as well as model selection and optimization) techniques for future work.

\noindent \underline{Decomposing pipelines.} 
We assume that it is possible to decompose a data-collection pipeline into self-contained tasks. 
However, such a decomposition may be cumbersome for complex learning problems like Puffer~\cite{puffer} that require closer service integration.

\noindent \underline{Decoupling pipelines from infrastructures.} 
We assume that it is possible to decouple the data-collection intents from actual infrastructure-specific mechanisms. 
However, realizing this may be difficult, especially for experiments where the data-collection tasks are heavily intertwined with a specific attribute of the data-collection node. 
For example, some IoT security experiments~\cite{unsw} require running the data-collection pipeline on specific devices with integrated firmware and pre-defined implementations of closed-source services, which cannot be easily supported by \system.

\noindent \underline{Programming overheads.} Our approach requires experimenters to express new data-collection tasks that are not yet presented in \system's library. 
Though this effort will amortize over time, it will only materialize if we succeed in building and incentivizing a broad user community for the proposed platform. 
Here, we take a first step and make a case for a holistic communal effort to address the data quality and model generalizability issues that have impeded the use of ML-based network security solutions in practice to date. 

\smartparagraph{Limitations of the prototype implementation.}

\noindent\underline{Data-collection nodes.} 
Our current prototype only supports Linux- or Windows-based nodes, optionally with Docker support to enable full platform capabilities (such as Docker container environments).
This restriction is reasonable because of the widespread support for Docker-based containers in current data-collection infrastructures~\cite{chiedge, edgenet} and a growing trend to manage Docker-based infrastructures~\cite{k8s,balena}. 
In future work, we plan to extend support to other computing environments, such as OpenWRT routers and PISA switches, which do not natively support Python or Docker. 
Currently, such extensions are possible using the sidecar model~\cite{sidecar}, which allows the configuration of nodes without Python support through Python-based APIs, such as P4-runtime~\cite{p4runtime}.

\noindent \underline{\rev{Potential subjectivity and biases.}} 
\rev{
Applying our proposed closed-loop ML pipeline involves the use of domain experts who themselves can be a source of possible biases or can make subjective decisions. One immediate solution to address this problem is to rely on multiple experts for cross-validation of explanations and decisions regarding data collection. For a more long-term solution, we envision the development of quantitative methods (e.g., metrics for evaluating explanation fidelity~\cite{guo2018lemna}) that will facilitate the detection of possible shortcuts or other types of inductive biases.}

\rev{
As far as other bias-related issues are concerned, we are already using a validation set for parameter selection to reduce parameter bias, and our method naturally helps avoid data snooping because it supports collecting data for different tasks and from different network environments at different times and allows for periodically examining and (if necessary) updating trained models.
}

\noindent \underline{\rev{Manual effort.}} 
\rev{A concerning side effect of using domain experts as part of our closed-loop ML pipeline is the manual effort it entails. While this makes the current version of our new pipeline inherently semi-automatic, future development of quantitative methods for detecting and possibly eliminating different types of inductive biases promises to reduce the manual effort required and make the pipeline more automatic. The development of such methods could potentially also benefit from advances in how AI can be utilized for examining model explanations and making model modification suggestions, but such issues are beyond the scope of this paper.}

\section{Related Work}
\label{sec:related}

\smartparagraph{Alternative approaches for our designs.}
In principle, it is possible to use existing tools and frameworks to realize the  ``thin waist" we implemented for data collection, but doing so while achieving
\system's level of abstraction, extensibility, fidelity, and scalability poses significant challenges (See \autoref{detailed_comparison_section} for details). 
For example, one possibility is to disaggregate pipelines into tasks with existing \textit{workflow-management platforms}, such as Airflow~\cite{airflow} or others~\cite{snakemake,luigi,dagster}.
However, there is often no explicit support to map these pipelines to specific data-collection nodes and instantiate multiple copies of tasks -- limiting data-collection experiments' flexibility. 
Existing \textit{CI/CD systems} (e.g., Jenkins~\cite{jenkins} and others~\cite{gitlabcicd, github_actions} allow explicit mapping of pipelines to nodes but typically require specific infrastructure access and configuration, limiting the desired extensibility and fidelity. Besides, they do not optimize inter-task execution time, limiting their ability to scale the data collection scenarios.
Finally, one can also use different \textit{configuration} (e.g., SaltStack~\cite{saltproject}) or \textit{orchestration platforms} (e.g., Kubernetes~\cite{k8s}), and others~\cite{ansible, chef, puppet, vsphere}. However, these systems lack the desired extensibility and flexibility because, being tailor-made for orchestration, they only work for specific types of infrastructures and do not provide explicit support for the proposed pipelines and stages abstraction, limiting tasks and experiments' reusability.

\smartparagraph{Passive data augmentation.}
In computer vision, researchers synthesize novel training data by adding random Gaussian noise to training images~\cite{van2001art,shorten2019survey} or blurring, rotating, and flipping them. However, these methods are specific to images and can only rarely be applied beyond vision data. 
Recent studies propose more application-domain independent methods, such as mixup~\cite{zhang2017mixup} and SMOTE~\cite{smote,smotestudy}, which can be applied to networking data. 
However, as demonstrated in \autoref{sec:eval_iterative}, these methods have limited efficacy in networking applications due to the correctness of the augmented data. 
They also generate samples that are typically very similar to the given training data, thus limiting the examination of model generalizability. 
Another line of data augmentation methods generates adversarial samples by adding carefully crafted perturbations to training samples (e.g.,~\cite{goodfellow2014explaining,chen2021towards,rebuffi2021data}).
Since these perturbations are just noises with a Non-Gaussian distribution, they suffer from similar limitations as adding Gaussian noise.

\smartparagraph{Model-side efforts.}
Various model-side efforts have also been considered to improve model generalizability. 
In particular, (reinforcement learning-based) domain adaptation methods (e.g.,\cite{shankar2018generalizing,farahani2021brief}) maintain an ML model's efficacy across multiple domains. 
To generalize across different learning problems, existing research proposed 
multi-task learning~\cite{ruder2017overview,zhang2018overview}) and few-shot learning methods~\cite{rivero2017grassmannian,goodfellow2016deep}. 
Researchers have also developed advanced models to combat shortcuts~\cite{geirhos2020shortcut} or out-of-distribution (ood) issues~\cite{hendrycks2016baseline}, such as detecting oods with contrastive learning~\cite{yang2021cade}.
All the model-side efforts assume that the training data is fixed and already given. 
These techniques are orthogonal and complementary to our method, which focuses on improving datasets. 

\section{Conclusion}
\label{sec:conclusion}

In this paper, we present a novel closed-loop ML pipeline to curate high-quality datasets for developing generalizable ML-based solutions for network security problems. 
Our approach is based on a new data-collection method that leverages advances in explainable ML and emphasizes the need for a flexible ``in vivo" collection of training datasets. 
It takes inspiration from the classic ``hourglass'' abstraction, where the different learning problems make up the hourglass' top layer, and the different network environments constitute its bottom layer. 
We realize the ``thin waist" of this hourglass abstraction with a new data-collection platform, \system. 
In effect, for each learning problem, \system enables data collection in multiple network environments, and for each network environment, it facilitates data collection for multiple learning problems. 
Through extensive experiments that involve different network security problems and consider multiple network infrastructures, we demonstrate how \system, in conjunction with the use of explainable ML tools, simplifies data collection for different learning problems from diverse network environments, enables iterative data collection for advancing the development of generalizable ML models, and improves the reproducibility, reusability, and shareability of network security experiments.
\begin{acks}
We thank the ACM CCS reviewers for their constructive feedback. NSF Awards CNS-2003257, OAC-2126327, and OAC-2126281 supported this work.
\end{acks}
\bibliographystyle{abbrv} 
\bibliography{references}

\begin{thebibliography}{100}

\bibitem{airflow}
Apache airflow.
\newblock \url{https://airflow.apache.org}.

\bibitem{atlas}
A.~Alsaheel, Y.~Nan, S.~Ma, L.~Yu, G.~Walkup, Z.~B. Celik, X.~Zhang, and D.~Xu.
\newblock Atlas: A sequence-based learning approach for attack investigation.
\newblock In {\em USENIX Security}, 2021.

\bibitem{257178}
Anonymous, A.~A. Niaki, N.~P. Hoang, P.~Gill, and A.~Houmansadr.
\newblock Triplet censors: Demystifying great {Firewall{\textquoteright}s} {DNS} censorship behavior.
\newblock In {\em FOCI}, 2020.

\bibitem{ansible}
Ansible automation platform.
\newblock \url{https://www.ansible.com/}.

\bibitem{apachepath}
Apache2 2.4.49 - lfi \& rce exploit.
\newblock \url{https://github.com/thehackersbrain/CVE-2021-41773}.

\bibitem{apley2019visualizing}
D.~W. Apley and J.~Zhu.
\newblock Visualizing the effects of predictor variables in black box supervised learning models, 2019.

\bibitem{arik2020tabnet}
S.~O. Arik and T.~Pfister.
\newblock Tabnet: Attentive interpretable tabular learning, 2020.

\bibitem{dos}
D.~Arp, E.~Quiring, F.~Pendlebury, A.~Warnecke, F.~Pierazzi, C.~Wressnegger, L.~Cavallaro, and K.~Rieck.
\newblock Dos and don{\textquoteright}ts of machine learning in computer security.
\newblock In {\em USENIX Security}, 2022.

\bibitem{ripeatlas}
Ripe atlas.
\newblock \url{https://atlas.ripe.net/}.

\bibitem{fabric}
I.~Baldin, A.~Nikolich, J.~Griffioen, I.~I.~S. Monga, K.-C. Wang, T.~Lehman, and P.~Ruth.
\newblock Fabric: A national-scale programmable experimental network infrastructure.
\newblock {\em IEEE Internet Computing}, 2019.

\bibitem{balena}
balena - the complete iot management platform.
\newblock \url{https://www.balena.io/}.

\bibitem{Ballmann_2021}
B.~Ballmann.
\newblock {\em Understanding Network Hacks}.
\newblock Springer Berlin Heidelberg, 2021.

\bibitem{invariant}
K.~Bartos, M.~Sofka, and V.~Franc.
\newblock Optimized invariant representation of network traffic for detecting unseen malware variants.
\newblock In {\em USENIX Security}, 2016.

\bibitem{10.1145/3274770}
M.~Beck.
\newblock On the hourglass model.
\newblock {\em Commun. ACM}, 62(7):48–57, jun 2019.

\bibitem{pinot}
R.~Beltiukov, S.~Chandrasekaran, A.~Gupta, and W.~Willinger.
\newblock Pinot: Programmable infrastructure for networking.
\newblock In {\em ANRW}, 2023.

\bibitem{279954}
A.~Bhaskar and P.~Pearce.
\newblock Many roads lead to rome: How packet headers influence {DNS} censorship measurement.
\newblock In {\em USENIX Security}, 2022.

\bibitem{272232}
H.~Birge-Lee, L.~Wang, D.~McCarney, R.~Shoemaker, J.~Rexford, and P.~Mittal.
\newblock Experiences deploying {Multi-Vantage-Point} domain validation at let{\textquoteright}s encrypt.
\newblock In {\em USENIX Security}, 2021.

\bibitem{breiman2001random}
L.~Breiman.
\newblock Random forests.
\newblock {\em Machine learning}, 45:5--32, 2001.

\bibitem{netmicroscope}
F.~Bronzino, P.~Schmitt, S.~Ayoubi, G.~Martins, R.~Teixeira, and N.~Feamster.
\newblock Inferring streaming video quality from encrypted traffic: Practical models and deployment experience.
\newblock {\em POMACS}, 2019.

\bibitem{aws}
Cloud computing services - amazon web services.
\newblock \url{https://aws.amazon.com/}.

\bibitem{azure}
Cloud computing services - microsoft azure.
\newblock \url{https://azure.microsoft.com/}.

\bibitem{digitalocean}
Cloud computing services - digitalocean.
\newblock \url{https://www.digitalocean.com/}.

\bibitem{gcp}
Cloud computing services - google cloud.
\newblock \url{https://cloud.google.com/}.

\bibitem{chiedge}
Chi@edge.
\newblock \url{https://chameleoncloud.org/experiment/chiedge/}.

\bibitem{chatzoglou2022revisiting}
E.~Chatzoglou, V.~Kouliaridis, G.~Karopoulos, and G.~Kambourakis.
\newblock Revisiting quic attacks: A comprehensive review on quic security and a hands-on study.
\newblock {\em International Journal of Information Security}, 2022.

\bibitem{smote}
N.~V. Chawla, K.~W. Bowyer, L.~O. Hall, and W.~P. Kegelmeyer.
\newblock {SMOTE}: Synthetic minority over-sampling technique.
\newblock {\em JAIR}, 2002.

\bibitem{chef}
Chef infra.
\newblock \url{http://www.chef.io/chef/}.

\bibitem{chen2021towards}
Z.~Chen, Q.~Li, and Z.~Zhang.
\newblock Towards robust neural networks via close-loop control.
\newblock {\em arXiv preprint arXiv:2102.01862}, 2021.

\bibitem{277132}
G.~Cherubin, R.~Jansen, and C.~Troncoso.
\newblock Online website fingerprinting: Evaluating website fingerprinting attacks on tor in the real world.
\newblock In {\em USENIX Security}, 2022.

\bibitem{cic}
Canadian institute for cybersecurity datasets.
\newblock \url{https://www.unb.ca/cic/datasets/index.html}.

\bibitem{cicflowmeter}
Cicflowmeter-v4.0.
\newblock \url{https://github.com/ahlashkari/CICFlowMeter}.

\bibitem{cuzzocrea2017tor}
A.~Cuzzocrea, F.~Martinelli, F.~Mercaldo, and G.~Vercelli.
\newblock Tor traffic analysis and detection via machine learning techniques.
\newblock In {\em Big Data}, 2017.

\bibitem{dagster}
Dagster.
\newblock \url{https://dagster.io/}.

\bibitem{Damour2020}
A.~D'Amour, K.~Heller, D.~Moldovan, B.~Adlam, B.~Alipanahi, A.~Beutel, C.~Chen, et~al.
\newblock Underspecification presents challenges for credibility in modern machine learning.
\newblock {\em Journal of Machine Learning Research}, 2022.

\bibitem{darpa}
1998 darpa intrusion detection evaluation dataset.
\newblock \url{https://www.ll.mit.edu/r-d/datasets/1998-darpa-intrusion-detection-evaluation-dataset}.

\bibitem{docker}
Docker.
\newblock \url{https://www.docker.com/}.

\bibitem{10.1145/3548606.3560604}
P.~Dodia, M.~AlSabah, O.~Alrawi, and T.~Wang.
\newblock Exposing the rat in the tunnel: Using traffic analysis for tor-based malware detection.
\newblock In {\em CCS}, 2022.

\bibitem{DraperGil2016CharacterizationOE}
G.~Draper-Gil, A.~H. Lashkari, M.~S.~I. Mamun, and A.~A. Ghorbani.
\newblock Characterization of encrypted and vpn traffic using time-related features.
\newblock In {\em ICISSP}, 2016.

\bibitem{deeplog}
M.~Du, F.~Li, G.~Zheng, and V.~Srikumar.
\newblock Deeplog: Anomaly detection and diagnosis from system logs through deep learning.
\newblock In {\em CCS}, 2017.

\bibitem{DHOOGE2020102564}
L.~D’hooge, T.~Wauters, B.~Volckaert, and F.~{De Turck}.
\newblock Inter-dataset generalization strength of supervised machine learning methods for intrusion detection.
\newblock {\em Journal of Information Security and Applications}, 54:102564, 2020.

\bibitem{edgenet}
Edgenet.
\newblock \url{https://www.edge-net.org/}.

\bibitem{farahani2021brief}
A.~Farahani, S.~Voghoei, K.~Rasheed, and H.~R. Arabnia.
\newblock A brief review of domain adaptation.
\newblock In {\em Advances in Data Science and Information Engineering}, 2021.

\bibitem{10.1214/aos/1013203451}
J.~H. Friedman.
\newblock {Greedy function approximation: A gradient boosting machine.}
\newblock {\em The Annals of Statistics}, 2001.

\bibitem{geirhos2020shortcut}
R.~Geirhos, J.-H. Jacobsen, C.~Michaelis, R.~Zemel, W.~Brendel, M.~Bethge, and F.~A. Wichmann.
\newblock Shortcut learning in deep neural networks.
\newblock {\em Nature Machine Intelligence}, 2020.

\bibitem{gepperth2020survey}
A.~Gepperth and S.~Rieger.
\newblock A survey of machine learning applied to computer networks.
\newblock In {\em ESANN}, 2020.

\bibitem{github_actions}
Github actions.
\newblock \url{https://docs.github.com/en/actions}.

\bibitem{gitlabcicd}
Gitlab ci/cd.
\newblock \url{https://docs.gitlab.com/ee/ci/}.

\bibitem{goodfellow2016deep}
I.~Goodfellow, Y.~Bengio, and A.~Courville.
\newblock {\em Deep learning}.
\newblock MIT press, 2016.

\bibitem{goodfellow2014explaining}
I.~J. Goodfellow, J.~Shlens, and C.~Szegedy.
\newblock Explaining and harnessing adversarial examples.
\newblock {\em arXiv preprint arXiv:1412.6572}, 2014.

\bibitem{10.1145/3523230.3523232}
M.~Gouel, K.~Vermeulen, M.~Mouchet, J.~P. Rohrer, O.~Fourmaux, and T.~Friedman.
\newblock Zeph iris map the internet: A resilient reinforcement learning approach to distributed ip route tracing.
\newblock {\em SIGCOMM Computer Communication Review}, 2022.

\bibitem{grinsztajn2022treebased}
L.~Grinsztajn, E.~Oyallon, and G.~Varoquaux.
\newblock Why do tree-based models still outperform deep learning on tabular data?, 2022.

\bibitem{guo2018lemna}
W.~Guo, D.~Mu, J.~Xu, P.~Su, G.~Wang, and X.~Xing.
\newblock Lemna: Explaining deep learning based security applications.
\newblock In {\em CCS}, 2018.

\bibitem{GUPTA2019466}
S.~Gupta and A.~Gupta.
\newblock Dealing with noise problem in machine learning data-sets: A systematic review.
\newblock {\em Procedia Computer Science}, 2019.

\bibitem{requet}
C.~Gutterman, K.~Guo, S.~Arora, T.~Gilliland, X.~Wang, L.~Wu, E.~Katz-Bassett, and G.~Zussman.
\newblock Requet: Real-time qoe metric detection for encrypted youtube traffic.
\newblock {\em ACM Transactions on MCCA}, 2020.

\bibitem{281442}
M.~Harrity, K.~Bock, F.~Sell, and D.~Levin.
\newblock {GET} /out: Automated discovery of {Application-Layer} censorship evasion strategies.
\newblock In {\em USENIX Security}, 2022.

\bibitem{heartbleed}
Heartbleed.
\newblock \url{https://gist.github.com/eelsivart/10174134}.

\bibitem{hendrycks2016baseline}
D.~Hendrycks and K.~Gimpel.
\newblock A baseline for detecting misclassified and out-of-distribution examples in neural networks.
\newblock {\em arXiv:1610.02136}, 2016.

\bibitem{Holland_2021}
J.~Holland, P.~Schmitt, N.~Feamster, and P.~Mittal.
\newblock New directions in automated traffic analysis.
\newblock In {\em CCS}, 2021.

\bibitem{hydra}
Hydra.
\newblock \url{https://github.com/vanhauser-thc/thc-hydra}.

\bibitem{trustee}
A.~S. Jacobs, R.~Beltiukov, W.~Willinger, R.~A. Ferreira, A.~Gupta, and L.~Z. Granville.
\newblock Ai/ml for network security: The emperor has no clothes.
\newblock In {\em CCS}, 2022.

\bibitem{jenkins}
Jenkins.
\newblock \url{https://www.jenkins.io/}.

\bibitem{jordaney2017transcend}
R.~Jordaney, K.~Sharad, S.~K. Dash, Z.~Wang, D.~Papini, I.~Nouretdinov, and L.~Cavallaro.
\newblock Transcend: Detecting concept drift in malware classification models.
\newblock In {\em USENIX Security}, 2017.

\bibitem{smotestudy}
G.~Kovács.
\newblock An empirical comparison and evaluation of minority oversampling techniques on a large number of imbalanced datasets.
\newblock {\em ASC}, 2019.

\bibitem{k8s}
Kubernetes - production-grade container orchestraction.
\newblock \url{https://kubernetes.io/}.

\bibitem{symprod}
I.~Kunakorntum, W.~Hinthong, and P.~Phunchongharn.
\newblock A synthetic minority based on probabilistic distribution (symprod) oversampling for imbalanced datasets.
\newblock {\em IEEE Access}, 2020.

\bibitem{mininet}
B.~Lantz, B.~Heller, and N.~McKeown.
\newblock A network in a laptop: Rapid prototyping for software-defined networks.
\newblock In {\em SIGCOMM Workshop on Hot Topics in Networks}, New York, NY, USA, 2010. Association for Computing Machinery.

\bibitem{log4j}
log4j-scan.
\newblock \url{https://github.com/fullhunt/log4j-scan}.

\bibitem{Lu_2018}
J.~Lu, A.~Liu, F.~Dong, F.~Gu, J.~Gama, and G.~Zhang.
\newblock Learning under concept drift: A review.
\newblock {\em {IEEE} Transactions on Knowledge and Data Engineering}, 2018.

\bibitem{luigi}
Luigi.
\newblock \url{https://github.com/spotify/luigi}.

\bibitem{shap}
S.~M. Lundberg and S.-I. Lee.
\newblock A unified approach to interpreting model predictions.
\newblock In {\em NeurIPS}. 2017.

\bibitem{maharana2022review}
K.~Maharana, S.~Mondal, and B.~Nemade.
\newblock A review: Data pre-processing and data augmentation techniques.
\newblock {\em Global Transitions Proceedings}, 2022.

\bibitem{memory-profiler}
memory-profiler.
\newblock \url{https://pypi.org/project/memory-profiler/}.

\bibitem{kitsune}
Y.~Mirsky, T.~Doitshman, Y.~Elovici, and A.~Shabtai.
\newblock Kitsune: An ensemble of autoencoders for online network intrusion detection.
\newblock In {\em NDSS}, 2018.

\bibitem{snakemake}
F.~Molder, K.~Jablonski, B.~Letcher, M.~Hall, C.~Tomkins-Tinch, V.~Sochat, J.~Forster, S.~Lee, S.~Twardziok, A.~Kanitz, A.~Wilm, M.~Holtgrewe, S.~Rahmann, S.~Nahnsen, and J.~Koster.
\newblock Sustainable data analysis with snakemake.
\newblock {\em F1000Research}, 2021.

\bibitem{molnar2020interpretable}
C.~Molnar.
\newblock {\em Interpretable machine learning}.
\newblock Lulu. com, 2020.

\bibitem{natekin2013gradient}
A.~Natekin and A.~Knoll.
\newblock Gradient boosting machines, a tutorial.
\newblock {\em Frontiers in neurorobotics}, 7:21, 2013.

\bibitem{mahimahi}
R.~Netravali, A.~Sivaraman, S.~Das, A.~Goyal, K.~Winstein, J.~Mickens, and H.~Balakrishnan.
\newblock Mahimahi: Accurate {Record-and-Replay} for {HTTP}.
\newblock In {\em USENIX ATC}, 2015.

\bibitem{netrics}
Netrics.
\newblock \url{https://github.com/chicago-cdac/nm-exp-active-netrics}.

\bibitem{p181system}
System code of netunicorn.
\newblock \url{https://github.com/netunicorn/netunicorn}.

\bibitem{p181library}
Library of tasks for netunicorn.
\newblock \url{https://github.com/netunicorn/netunicorn-library}.

\bibitem{p181suppl}
Supplementary materials for netunicorn paper.
\newblock \url{https://github.com/netunicorn/netunicorn-search}.

\bibitem{nori2019interpretml}
H.~Nori, S.~Jenkins, P.~Koch, and R.~Caruana.
\newblock Interpretml: A unified framework for machine learning interpretability.
\newblock {\em arXiv preprint arXiv:1909.09223}, 2019.

\bibitem{ns3}
ns-3 | a discrete-event network simulator for internet systems.
\newblock \url{https://www.nsnam.org/}.

\bibitem{p0f}
p0f v3 (version 3.09b).
\newblock \url{https://lcamtuf.coredump.cx/p0f3/}.

\bibitem{p4runtime}
P4runtime specification.
\newblock \url{https://p4.org/p4-spec/p4runtime/main/P4Runtime-Spec.html}.

\bibitem{patator}
Patator.
\newblock \url{https://github.com/lanjelot/patator}.

\bibitem{pawr}
Platforms for advanced wireless research.
\newblock \url{https://advancedwireless.org/}.

\bibitem{PETCH2022204}
J.~Petch, S.~Di, and W.~Nelson.
\newblock Opening the black box: The promise and limitations of explainable machine learning in cardiology.
\newblock {\em Canadian Journal of Cardiology}, 2022.

\bibitem{puppet}
Puppet.
\newblock \url{https://puppet.com/}.

\bibitem{network_attacks}
Python network attacks.
\newblock \url{https://github.com/PacktPublishing/Basic-and-low-level-Python-Network-Attacks}.

\bibitem{quinonero2008dataset}
J.~Quinonero-Candela, M.~Sugiyama, A.~Schwaighofer, and N.~D. Lawrence.
\newblock {\em Dataset shift in machine learning}.
\newblock Mit Press, 2008.

\bibitem{rebuffi2021data}
S.-A. Rebuffi, S.~Gowal, D.~A. Calian, F.~Stimberg, O.~Wiles, and T.~A. Mann.
\newblock Data augmentation can improve robustness.
\newblock In {\em NeurIPS}, 2021.

\bibitem{lime}
M.~T. Ribeiro, S.~Singh, and C.~Guestrin.
\newblock "why should i trust you?": Explaining the predictions of any classifier.
\newblock In {\em Proceedings of the 22nd ACM SIGKDD International Conference on Knowledge Discovery and Data Mining}, KDD '16, page 1135–1144, New York, NY, USA, 2016. Association for Computing Machinery.

\bibitem{richards2015software}
M.~Richards.
\newblock {\em Software Architecture Patterns: Understanding Common Architecture Patterns and when to Use Them}.
\newblock O'Reilly Media, 2015.

\bibitem{rivero2017grassmannian}
J.~Rivero, B.~Ribeiro, N.~Chen, and F.~S. Leite.
\newblock A grassmannian approach to zero-shot learning for network intrusion detection.
\newblock In {\em ICONIP}, 2017.

\bibitem{ruder2017overview}
S.~Ruder.
\newblock An overview of multi-task learning in deep neural networks.
\newblock {\em arXiv preprint arXiv:1706.05098}, 2017.

\bibitem{saltproject}
Salt project.
\newblock \url{https://saltproject.io/}.

\bibitem{beautyburst}
R.~Schuster, V.~Shmatikov, and E.~Tromer.
\newblock Beauty and the burst: Remote identification of encrypted video streams.
\newblock In {\em USENIX Security}, 2017.

\bibitem{seclists}
Seclists.
\newblock \url{https://github.com/danielmiessler/SecLists}.

\bibitem{shankar2018generalizing}
S.~Shankar, V.~Piratla, S.~Chakrabarti, S.~Chaudhuri, P.~Jyothi, and S.~Sarawagi.
\newblock Generalizing across domains via cross-gradient training.
\newblock {\em arXiv preprint arXiv:1804.10745}, 2018.

\bibitem{cicids2017}
I.~Sharafaldin, A.~H. Lashkari, and A.~A. Ghorbani.
\newblock Toward generating a new intrusion detection dataset and intrusion traffic characterization.
\newblock In {\em International Conference on Information Systems Security and Privacy}, 2018.

\bibitem{https://doi.org/10.48550/arxiv.1911.01058}
S.~Shi, X.~Zhang, and W.~Fan.
\newblock Explaining the predictions of any image classifier via decision trees, 2019.

\bibitem{shorten2019survey}
C.~Shorten and T.~M. Khoshgoftaar.
\newblock A survey on image data augmentation for deep learning.
\newblock {\em Journal of big data}, 2019.

\bibitem{tabular22}
R.~Shwartz-Ziv and A.~Armon.
\newblock Tabular data: Deep learning is not all you need.
\newblock {\em Information Fusion}, 2022.

\bibitem{sidecar}
Sidecar.
\newblock \url{https://docs.microsoft.com/en-us/azure/architecture/patterns/sidecar}.

\bibitem{281438}
J.-P. Smith, L.~Dolfi, P.~Mittal, and A.~Perrig.
\newblock {QCSD}: A {QUIC} {Client-Side} {Website-Fingerprinting} defence framework.
\newblock In {\em USENIX Security}, 2022.

\bibitem{unsw}
Unsw datasets.
\newblock \url{https://iotanalytics.unsw.edu.au/}.

\bibitem{van2001art}
D.~A. Van~Dyk and X.-L. Meng.
\newblock The art of data augmentation.
\newblock {\em Journal of Computational and Graphical Statistics}, 2001.

\bibitem{Vasi_2022}
M.~Vasi{\'{c}}, A.~Petrovi{\'{c}}, K.~Wang, M.~Nikoli{\'{c}}, R.~Singh, and S.~Khurshid.
\newblock {MoËT}: Mixture of expert trees and its application to verifiable reinforcement learning.
\newblock {\em Neural Networks}, 151:34--47, jul 2022.

\bibitem{vsphere}
Vmware vsphere.
\newblock \url{https://www.vmware.com/products/vsphere.html}.

\bibitem{webdav}
Web distributed authoring and versioning (webdav) ordered collections protocol.
\newblock \url{https://www.rfc-editor.org/rfc/rfc3648.html}.

\bibitem{xnids}
F.~Wei, H.~Li, Z.~Zhao, and H.~Hu.
\newblock Xnids: Explaining deep learning-based network intrusion detection systems for active intrusion responses.
\newblock In {\em Security}, 2023.

\bibitem{927}
Overview of competitive standards.
\newblock \url{https://xkcd.com/927/}.

\bibitem{puffer}
F.~Y. Yan, H.~Ayers, C.~Zhu, S.~Fouladi, J.~Hong, K.~Zhang, P.~Levis, and K.~Winstein.
\newblock Learning in situ: a randomized experiment in video streaming.
\newblock In {\em NSDI}, 2020.

\bibitem{pantheon}
F.~Y. Yan, J.~Ma, G.~D. Hill, D.~Raghavan, R.~S. Wahby, P.~Levis, and K.~Winstein.
\newblock Pantheon: the training ground for internet congestion-control research.
\newblock In {\em USENIX ATC}, 2018.

\bibitem{yang2021cade}
L.~Yang, W.~Guo, Q.~Hao, A.~Ciptadi, A.~Ahmadzadeh, X.~Xing, and G.~Wang.
\newblock $\{$CADE$\}$: Detecting and explaining concept drift samples for security applications.
\newblock In {\em USENIX Security}, 2021.

\bibitem{zhang2017mixup}
H.~Zhang, M.~Cisse, Y.~N. Dauphin, and D.~Lopez-Paz.
\newblock mixup: Beyond empirical risk minimization.
\newblock {\em arXiv preprint arXiv:1710.09412}, 2017.

\bibitem{zhang2018overview}
Y.~Zhang and Q.~Yang.
\newblock An overview of multi-task learning.
\newblock {\em NSR}, 2018.

\bibitem{zhou2019evaluation}
Q.~Zhou and D.~Pezaros.
\newblock Evaluation of machine learning classifiers for zero-day intrusion detection--an analysis on cic-aws-2018 dataset.
\newblock {\em arXiv preprint arXiv:1905.03685}, 2019.

\end{thebibliography}

\appendix
\section{Validating Let's Encrypt challenges from multiple vantage points.}
\label{appendix:letsencrypt}
In this scenario, we consider the task of domain name validation via the ACME challenge by Let's Encrypt. Recent papers~\cite{272232} argue for the importance of using multiple vantage points for performing this task, 
where the vantage point should be both geographically and logically dispersed across different networks to avoid BGP attacks and prevent the validation of malicious requests. 

We used \system to implemented the DNS-01 and HTTP-01 validation protocols for the ACME challenge and to create an experiment with nodes in two different infrastructures (\university and multi-region Azure), effectively mimicking the multi-vantage point scenario from the original paper~\cite{272232}. We enhanced the experiment by supporting dynamic node selection, thus making possible BGP attacks more difficult due to \textit{a priori} unknown vantage point location. We expressed this experiment using only \textbf{14 LLoCs}, excluding challenge protocol implementation (see corresponding tasks in \autoref{appendix:tasks}).
\section{Additional iterative experiments.}
\label{additional_iterative}

\rev{In this Appendix, we describe two additional network security problems that could benefit from our proposed iterative approach. %
In each case, we include a description of the problem, 
describe the training data used by existing learning models, and discuss underspecification issues associated with these datasets. 
Next, we demonstrate how \system can be utilized to express data collection intents for the given problem, especially for the first problem that considers the widely-used CIC-IDS-2017 setup. 
Finally, we explain how \system can be leveraged to refine the data-collection experiment and collect new data to address the previously reported underspecification issues.}

\subsection{\rev{Heartbleed detection.}}
\rev{This scenario concerns the Heartbleed detection problem~\cite{heartbleed} and has been previously studied in the context of the CIC-IDS-2017 dataset~\cite{cicids2017}. A Heartbleed attack is a specifically constructed network packet that tries to use a heartbeat vulnerability in the OpenSSL library to obtain random memory bytes from a target server.}

\rev{We consider the Heartbleed attack data that is part of the CIC-IDS-2017 dataset. The data is given in the form of CICFlowMeter features that we also used in the \autoref{sec:eval_iterative}. These features describe different flow statistics, such as packet inter-arrival time (mean, min, max, std), packet size (mean, min, max. std), and others.}

\rev{Considering the CIC-IDS-2017 data to represent the dataset for the initial iteration of our iterative data-collection approach, we can use explainable ML techniques as part of our newly proposed closed-loop ML pipeline to explore the data for possible shortcuts and other types of underspecification issues. Using Trustee, the authors of~\cite{trustee} showed that for the considered dataset, it was possible to detect all Heartbleed examples by simply checking the "Bwd Packet Length Max" feature. Since in the Heartbleed case, attackers try to collect as much of the target's memory as possible to extract potentially valuable data from the target, many Heartbleed attack patterns require a server to return packets with a big payload, which is easily detectable in the resulting dataset.}

\rev{Since for an arbitrary server hosting web pages, backward packet size typically varies (e.g., small for simple requests, large for returning binary objects), we consider the exclusive use of the "Bwd Packet Length Max" feature to identify Heartbleed attacks to be an instance of shortcut learning. To mitigate this shortcut, we can leverage \system and implement and perform various realistic benign traffic pattern tasks (e.g., requesting large files, streaming) that result in variable-sized backward packets. This change in how benign traffic is generated will for all practical purposes eliminate the observed dependency on this single feature for this attack, effectively eliminating the root cause in the data that was responsible for the identified shortcut.}

\rev{After eliminating the noted data issue and using \system to collect a new dataset (with benign traffic generated as described above), we can again apply explainable ML techniques to investigate the resulting data for possible data issues. In fact, as shown in~\cite{trustee}, for black-box models trained with this new dataset, Trustee identifies "Bwd IAT Total” (Backward Total Inter-Arrival Time) as the sole feature capable of perfectly separating Heartbleed attacks from benign traffic. The reason for this is an attack implementation bug that prevents the closing of TCP sessions between successive attacks. As a result, single TCP connections stay open for unusually long periods of time, and this behavior allows for easy and accurate identification of Heartbleed attacks in the collected data.}

\rev{However, in real-world scenarios, the Heartbleed connection is usually closed after the attack and reopened when a new attack is initiated. As a result, we consider the sole use of the "Bwd IAT Total” feature to define yet another shortcut, this time caused by a Heartbleed attack implementation flaw. Having recognized and identified this issue with the collected data, we can again use our new closed-loop ML pipeline to first modify the source code of the Heartbleed attack so as to avoid the noted original implementation bug, then redeploy the attacking pipeline to the same nodes as in the original scenario, and finally collect a new dataset. Note that this last dataset is of higher quality than the original CIC-IDS-2017 dataset in the sense that the root causes for both identified shortcuts are no longer present. As a result, the described approach results in datasets that improve the generalizability of ML models that utilize these data for training. Importantly, the thus-trained models have a better chance to perform well in different network scenarios.}

\subsection{\rev{OS Fingerprinting.}}
\rev{This scenario considers the Operating System Fingerprinting learning problem described in the nPrint paper~\cite{Holland_2021}. Here, the problem is to use flow- and packet-level information (e.g., packet headers) to detect the operating system of the source of the network traffic flow. Existing tools such as p0f~\cite{p0f} deal with this problem by relying on different manual heuristics and packet analysis.}

\rev{We leverage the OS Fingerprinting training data that is part of the dataset published in the nPrint paper. This dataset contains PCAP files and OS source information for each flow. The data is represented as a nPrint vector that contains bits for the fields in each header of the first five packets in the flow.}

\rev{Considering this data to be the dataset for the initial iteration of our iterative data-collection approach, we can again use explainable ML techniques to identify the most important features that ML models trained with this data utilize as part of their decision-making. In fact, for this dataset, the authors of~\cite{trustee} showed that TTL (time-to-live) is the most important feature for accurately identifying OS types. This correlates with known default TTL values for different OSes (e.g., 64 and 128 for Linux and Windows, respectively). However, in the given dataset, Kali Linux is easily identified from among all other Linux systems due to the fact that it uses a lower TTL than the default value (i.e., 126 instead of 128).}

\rev{Upon closer inspection of how the nPrint data was collected, the observed difference in TTL values can be traced to the fact that Kali Linux was only used for attacking machines, all of which were located ``outside" of the network (where the benign traffic was generated) and had exactly two routers between them and the traffic collection point. Given that this information is not related to Kali Linux-specific aspects or properties but derives exclusively from the considered network configuration and the particular data collection setup, we consider the sole use of the TTL feature for OS fingerprinting to be an instance of shortcut learning.}

\rev{To eliminate this issue with the data, we can use \system to redeploy attacking and benign pipelines to different machines so as to ensure more diversity in measured TTL values. Thus, after eliminating this way the root cause for the identified shortcut in the original data, we can leverage \system to recollect data and then use the newly obtained data for model training. This will result in trained models for the OS fingerprinting problem that are better able to generalize than the ones trained with the original nPrint data and are therefore expected to have improved performance when deployed in real-world environments.}

\section{Expanding Iterative Collection}
We also consider an expanded version of the experiment conducted in \autoref{sec:eval_iterative}. In this version, we use the \university environment for training and both the campus-cloud and multi-cloud environments for testing. In addition, instead of having a fixed testing dataset, we collect testing datasets using the same experiment modifications as for training infrastructure, mitigating the possible distribution difference between training and testing data. Results are presented in the \autoref{tab:iterative_expanded} and align with the original experiment in \autoref{sec:eval_iterative}, showing improved model generalizability with each iteration.

\begin{table*}[t]
\centering
\caption{{Number of LLoC changes, data points, and F1 scores across different environments and iterations.}
\label{tab:iterative_expanded}
}
\resizebox{\textwidth}{!}{%
\begin{tabular}{c|c|cc|c|cc|c|cc}
& \multicolumn{3}{c}{Initial setup (iteration \#0)} & \multicolumn{3}{c}{Iteration~1 } & \multicolumn{3}{c}{Iteration~2} \\
\hline
LLoCs & \multicolumn{3}{c}{\textbf{80}} & \multicolumn{3}{c}{\textbf{+10}} & \multicolumn{3}{c}{\textbf{+20}} \\
\hline
& UCSB & UCSB-cloud & multi-cloud
& UCSB & UCSB-cloud & multi-cloud 
& UCSB & UCSB-cloud & multi-cloud 
\\
\hline
Data points &  [5.6~k, 1~k]     &  [0.5~k, 0.3~k]  &  [5.6~k, 0.6~k]      
&  [13.6~k, 1.8~k]     &  [10.5~k, 16~k]     &  [11.2~k, 2.0~k]    
&  [91~k, 59~k]     &  [178.8~k, 106.9~k]     &  [133.8~k, 49.8~k]      \\
\hline
MLP & 1.0 & 0.59 & 0.66 & 0.97 (-0.03) & 0.82 (+0.23) & 0.72 (+0.06) & 0.88 (-0.11) & \textbf{0.93 (+0.11)} & \textbf{0.94 (+0.22)} \\
GB  & 1.0 & 0.32 & 0.71 & 1.0 (+0.00) & 0.78 (+0.46) & 0.67 (-0.04) & 0.92 (-0.08) & \textbf{0.94 (+0.16)} & \textbf{0.92 (+0.25)}           \\
RF  & 1.0 & 0.42 & 0.67 & 1.0 (+0.00) & 0.57 (+0.15) & 0.75 (+0.08) & 0.97 (-0.03) & \textbf{0.93 (+0.36)} & \textbf{0.93 (+0.18)}
\end{tabular}%
}
\end{table*}
\begin{table*}[t]
\caption{\system's API.}
\label{tab-api}
\begin{center}

    \resizebox{.85\linewidth}{!}{%
        \begin{tabular}{|l|l|l|}
        \hline
        \textbf{Object} & \textbf{Operations} & \textbf{Description} \\
        \hline
        {\tt Task} &  {\tt run()} & Entry point for task execution code \\
        \hline
        {\tt Pipeline} & {\tt then([tasks])} & Create a new stage of execution for the pipeline and add tasks to it  \\
        \hline
        \multirow{2}{*}{{\tt Nodes}} & {\tt filter(pred)}  & Filter nodes based on given predicate \\
        & {\tt take(N)} & Return no more than N nodes with filters applied \\
        \hline 
        \multirow{1}{*}{{\tt Experiment}} & {\tt map(pipeline, hosts)} & Assign a pipeline to a host(s) and choose appropriate task implementation \\ \hline
        \multirow{4}{*}{{\tt Client}}
        & {\tt deploy()} & Start environment compilation and distribution of the experiment \\
        & {\tt execute()} & Start execution of the deployed experiment \\
        & {\tt status()} & Returns status of the experiment (ready, running, finished, etc.) \\
        \hline
        \end{tabular}
}
\end{center}

\end{table*}
\section{Implemented connectors}
\label{appendix:connectors}
As a part of the system development, we implemented a number of connectors to different infrastructures or deployment systems. Each of these connectors is configurable, complete, and publicly available at our GitHub organization. \autoref{table:deploymentsystems} provides a list of the connectors and corresponding logical lines of code for their implementation. We encourage other research groups and individuals to improve existing or create and publish new connectors for deployment systems and infrastructures we haven't covered yet.

\begin{table}[t]
\caption{Implemented connectors to different Deployment Systems and corresponding LLoCs.}
\label{table:deploymentsystems}
\resizebox{.6\linewidth}{!}{
\centering
\begin{tabular}{l|c}
\textbf{Deployment Systems} & LLoCs \\ \hline
SaltStack & 205 \\ 
Azure Container Instances & 138 \\
Local Docker containers & 163 \\
\rev{Containernet} & 242 \\
AWS Fargate & 179 \\
Kubernetes & 197 \\
SSH & 186 \\
\end{tabular}
}
\vspace{-10pt}
\end{table}
\section{Implemented Tasks Description}
\label{appendix:tasks}

We briefly describe the full list of tasks that we implemented for \system. For each task, we provide the task intent, the number of logical lines of code (LLoC) for standard task implementation, and the number of LLoC to implement a wrapper for \system. The results are provided in the \autoref{table:appendixtaskloc}.

\begin{table*}[ht]
\caption{Implemented tasks description and corresponding LLoC for task and wrapper implementation. Most of the \textit{wrapper} code is constant and repetitive and adds little actual overhead for the implementation.}
\label{table:appendixtaskloc}
\resizebox{\linewidth}{!}{
\begin{tabular}{|c|l|l|cc|c|}
\hline
 &
\textbf{Task} &
\textbf{Description} &
\textbf{Core} &
\textbf{Wrapper} &
\textbf{Total}\\ \hline
1 & {\tt DummyTask} & Empty task & 0 & 4 & 4 \\
2 & {\tt SleepTask} & Sleep for a given amount of seconds & 1 & 7 & 8 \\
3 & {\tt ShellCommand} & Executes a given command in the system shell & 1 & 6 & 7\\
4 & {\tt Ping} & Executes a \textit{ping} command to a target host & 65 & 22 & 87\\
5 & {\tt PortScan} & Check if a port on a remote host is open & 4 & 6 & 10\\
6 & {\tt ArpSpoof} & ARP poisoning attack~\cite{Ballmann_2021} & 13 & 11 & 24\\
7 & {\tt FakeMail} & Sends a mail with a fake sender via unprotected mail server~\cite{Ballmann_2021} & 8 & 9 & 17\\
8 & {\tt MACFlooder} & Floods the network with packets with random IP and MAC~\cite{Ballmann_2021} & 8 & 9 & 17 \\
9 & {\tt SlowLoris} & Slowloris DoS attack~\cite{network_attacks} & 72 & 12 & 84 \\
10 & {\tt SMBloris} & SMBloris attack~\cite{network_attacks} & 19 & 11 & 30 \\
11 & {\tt LANDAttack} & LAND attack in the network~\cite{network_attacks} & 13 & 11 & 24\\
12 & {\tt ICMPRedirection} & ICMP redirection attack~\cite{network_attacks} & 6 & 10 & 16\\
13 & {\tt Patator} & Patator~\cite{patator} HTTP endpoint Basic authorization bruteforce & 37 & 14 & 51 \\
14 & {\tt Hydra} & Hydra~\cite{hydra} HTTP endpoint bruteforce & 14 & 10 & 24 \\
15 & {\tt CVE20140160} & CVE-2014-0160 (Heartbleed)~\cite{heartbleed} vulnerability exploit & 74 & 32 & 106 \\
16 & {\tt CVE202141773} & CVE-2021-41773 (Apache 2.4.49 Path)~\cite{apachepath} vulnerability exploit & 7 & 7 & 14 \\
17 & {\tt CVE202144228} & CVE-2021-44228 (Log4J)~\cite{log4j} vulnerability exploit & 5 & 7 & 12 \\
18 & {\tt UploadToWebDav} & Uploads a given set of files to a WebDAV~\cite{webdav} server & 7 & 10 & 17 \\
19 & {\tt StartCapture, StopAllTCPDumps} & Start and stop of \textit{tcpdump} tool for capturing the network traffic & 7 & 10 & 17 \\
20 & {\tt YouTubeWatcher} & Implementation of headless video watcher for the YouTube website & 61 & 22 & 83 \\
21 & {\tt TwitchWatcher} & Implementation of headless video watcher for the Twitch website & 28 & 20 & 48 \\
22 & {\tt VimeoWatcher} & Implementation of headless video watcher for the Vimeo website & 48 & 22 & 70\\
23 & {\tt QoECollectionServer} & Implementation of a task for YouTube QoE statistics collection & 46 & 28 & 74 \\
24 & {\tt LetsEncryptDNS01Validation} & Implementation of DNS-01 challenge validation for Let's Encrypt & 11 & 9 & 20 \\
25 & {\tt LetsEncryptHTTP01Validation} & Implementation of HHTP-01 challenge validation for Let's Encrypt & 11 & 10 & 21 \\ \hline
& \textbf{Total} & & 562 & 313 & 875 \\
\hline
\end{tabular}
}
\end{table*}

\section{Scaling Data Collection}
\label{appendix:measurements}

We quantify how our design choices help reduce the computing and memory overheads incurred by \system's core and executor(s).

\smartparagraph{Executors.}
Recall that for each experiment, \system's {\tt mediation service} requests the {\tt connectivity-manager} to instantiate an executor for all the participating data-collection nodes.
Our goal is to quantify the executor's overhead for a (relatively) low-end data-collection node, i.e., a Raspberry Pi (RPi) 4B device at our \university infrastructure. 
To ensure that our measurements are not skewed by the nature of the data-collection tasks, processing stages, and pipelines, we created custom pipelines with varying numbers of tasks and stages for our evaluation.
Specifically, we evaluated four pipelines: (1) a short pipeline with one stage and one task, (2) a short pipeline with two stages and ten tasks per stage, (3) a long pipeline with 100 stages and one task per stage, and (4) a long pipeline with 100 stages and ten tasks per stage. 
Each task in all these pipelines sleeps for 5 seconds.

For each pipeline, we quantify the executor's computing overhead as the difference between the completion time for different tasks and processing stages and related sleep times.
We observe that the executor's average computing overhead is \textbf{1 second} per stage and \textbf{0.13 seconds} per task in all pipelines, including the overhead for process spawning, data serialization, and results collection.
We measure the executor's memory overhead using a Python-based tool, \textit{memory-profiler}~\cite{memory-profiler}.
We observe that the executor's total memory overhead is \textbf{20.2 MB}, with the pipeline size from 1 to 19 KB. 
These results show that the executor's low computing and memory overheads will not negatively impact the pipeline's completion time or data quality, even for low-end devices like RPis.

\begin{table}[t]
\caption{Wall-time (seconds) overhead of different stages of experiments, required for services interaction. Due to the specific nature of ACI, the steps for image distribution and execution have been merged, as indicated by the underlined text in the table.}
\label{tab:platformtimeoverhead}
\centering
\resizebox{0.75\linewidth}{!}{
\begin{tabular}{c|ccc|ccc}
 &
\multicolumn{3}{c|}{\university} &
\multicolumn{3}{c}{ACI} \\ 
Nodes \# & 1 & 10 & 20 & 1 & 10 & 20 \\ 
\hline
Deployment & 3 & 4 & 3 & 5 & 4 & 5 \\ %
Execution & 4 & 13 & 19 & \underline{\textit{31}} & \underline{\textit{47}} & \underline{\textit{49}}  \\
\end{tabular}
}
\end{table}

\smartparagraph{\system's core.}
To quantify overheads incurred by \system's core, we use the data-collection experiment for the bruteforce attack detection problem.
For this experiment, we collect data from two infrastructures: \university (with RPis) and Azure Container Instances (ACI) (with AMD64-based Linux containers).
For both infrastructures, we expressed an experiment that uses a different number of data-collection nodes: 1, 10, and 20. For both of these infrastructures, it is possible to configure the computing environment locally and ship the configured docker image to the data-collection nodes. 

We report two metrics to quantify the computing overheads: \textit{deployment overhead} and \textit{execution overhead}.
\textit{Deployment overhead} measures the wall-clock time between the instance when an experiment is submitted to the time when it is ready for execution minus the time it takes to configure the docker image and distribute the instructions to the respective data-collection nodes. 
\textit{Execution overhead} measures the wall-clock time between the start and end times of an experiment minus the wall-clock time for individual tasks. 
Please refer to \autoref{appendix:dockerstages} for more details about an experiment's lifecycle in \system for docker-based infrastructures.

\autoref{tab:platformtimeoverhead} shows the wall-clock overhead for both stages. Note that we report the image distribution time as part of the execution overhead for the Azure Container Instances -- due to available operations in Azure Cloud SDK, it is impossible to separate these stages. 
We also measured the total memory overhead of the platform on our servers (a single SuperMicro server platform with AMD64 architecture and Ubuntu 22.04). 
All services (\textbf{6} in total) were implemented using Python 3.11, deployed in Docker containers, and in total consumed \textbf{240 MB}. 
In addition, the platform requires a PostgreSQL database for storing states, pipelines, and results, and optionally a private docker repository for image storage.

In summary, this evaluation shows the memory and computing efficiency of \system's core and executor(s)---demonstrating its ability to scale data-collection in realistic settings.

\section{Experiment Preparation and Execution Breakdown}
\label{appendix:dockerstages}
We provide a breakdown of a typical experiment preparation and execution with a Docker environment:
\begin{enumerate}
    \item User defines or imports tasks that should be executed on the nodes and combines them into pipelines.
    \item User requests a node pool from the platform, defines an experiment by assigning pipelines to nodes, and submits it to the \system.
    \item Platform analyzes the assignment of pipelines and defines Docker images to compile. This stage could be skipped if for all pipelines a custom prebuilt image is provided.
    \item \system's service compiles requested images and uploads them to a repository.
    \item \system requests connector to upload images to the nodes. This stage could be skipped if custom images were provided and they are already presented on the target nodes.
    \item \system marks the experiment as \textit{READY}.
    \item User requests the platform to start a ready experiment.
    \item \system requests connector to distribute the start command to all ready nodes participating in the experiment.
    \item Each node starts the container with an executor which executes the tasks and reports results back to the platform.
    \item The platform awaits for all nodes to report the results or time out, and then sets the experiment status to \textit{FINISHED}.
\end{enumerate}

\begin{table*}[!htbp]
\caption{A comparison between Workflow Management Platforms (WMP), Orchestration Platforms (OP), Continuous Integration / Continuous Deployment tools (CI/CD), and \system. In the table, + stands for \textit{mainly provided by a majority of tools}, - for \textit{unsupported by the majority of tools}, -/+ represents the \textit{mixed support}, and ? is used for \system to represent extensible features to be implemented in near future.}
\label{detailed_comparison_table}
\resizebox{\linewidth}{!}{
\begin{tabular}{|l|l|c|c|c|c|}
\hline
\textbf{Requirement} &
\textbf{Feature name} &
  \textbf{WMP} &    %
  \textbf{OP} &  %
  \textbf{CI/CD} &
  \textbf{\system} \\ \hline
\multirow{4}{*}{Extensibility} & Pipeline and Task abstractions            & + & - & +   & + \\
& Complex directed acyclic graphs (conditions, loops)           & + & –/+ & –/+ & ? \\
& Explicit node selection mechanisms        & -  & +   & +   & + \\
& Different executor architecture (Linux, Windows, OpenWRT, etc.) & -/+ & -/+ & -/+ & + \\ \cline{1-1}

\multirow{3}{*}{Scalability} & Pipeline execution synchronization       & + & –/+ & -   & + \\
& Low runtime execution overhead            & – & + & -   & + \\

 & Multiple node environments (shells, containers, VMs) & + & – & +   & + \\ \cline{1-1}

\multirow{2}{*}{Other}
& Cross-instance experiment synchronization & – & –   & –   & ? \\
& Data analytics platforms integration      & + & –   & –   & ? \\
\hline
\end{tabular}
}
\end{table*}
\section{Comparison with Existing Classes of Tools.}
\label{detailed_comparison_section}

Here we provide a more detailed comparison of \system with existing classes of tools suitable for data collection purposes in the networking area~\cite{927}, mentioned in \autoref{sec:related}.
We consider three main classes of tools that can enable data collection for our scenarios and provide a combined description of their differences from our system in \autoref{detailed_comparison_table}.

\smartparagraph{Workflow management platforms.} 
These solutions are designed to define and execute a data processing pipeline using one of the available platforms. 
Typical examples of such systems are Airflow~\cite{airflow}, SnakeMake~\cite{snakemake}, Luigi~\cite{luigi}, Dagster~\cite{dagster}, and others. 
Unfortunately, these systems do not always provide convenient ways of selecting nodes for code execution (relying on affinity settings, like Airflow Kubernetes operator or similar), which is critical for network experiments for precise data collection control. 
They also rarely try to minimize system overhead (especially between task execution) and require nodes to have a constant stable connection to the platform, which is not always available in our scenarios (e.g., nodes could be situated in remote locations with intermittent network connectivity).

\smartparagraph{Orchestration platforms.} 
Such systems are usually used to change the configuration of controlled nodes (servers, laptops, etc.) or deploy containers or virtual machines to particular nodes. 
Common examples of these systems are Ansible~\cite{ansible}, SaltStack~\cite{saltproject}, Chef~\cite{chef}, Puppet~\cite{puppet}, and Kubernetes~\cite{k8s}, VMware vSphere~\cite{vsphere} for containers and VMs deployment.
These systems typically need a specific infrastructure setup and administration, which requires root access to nodes. 
They are challenging to integrate with or run alongside other systems, limiting their implementation in other infrastructures. 
These systems' pipelines (playbooks) are often customized with unique information about certain nodes, complicating mapping them to other nodes or infrastructures.

\smartparagraph{Continuous integration and continuous delivery tools.} 
These tools provide a way to execute a set of instructions on specified nodes, usually for application development automation or deployment. 
The most popular examples of such systems are Jenkins~\cite{jenkins}, Gitlab CI/CD~\cite{gitlabcicd}, and Github Actions~\cite{github_actions}. 
These tools can be adjusted for data collection. 
Still, they do not optimize important data generation properties (such as overhead between tasks), use declarative language for configuration, do not separate deployment and execution of pipelines, or restrict the scalability of solutions (e.g., GitHub Actions Free plan supports only 20 parallel jobs, and only up to 180 parallel jobs in GitHub Enterprise).

\smartparagraph{Specialized data-collection platforms and infrastructures.}
This category includes platforms designed for specific (often community-based) data-collection experiments. 
Popular examples include platforms such as RIPE Atlas~\cite{ripeatlas}, Puffer experiment~\cite{puffer}, Netrics~\cite{netrics}, etc. 
Unfortunately, these platforms cannot be easily extended to support data collection for multiple learning problems from one or more network environments. 

\section{Source Code and Supplementary Materials}
\label{appendix:opensource}
In this section, we describe the \system repositories and their purpose.

\smartparagraph{\system's code}. The system's code is available in this repository: \url{https://github.com/g4allthewaydown/paper-181-system}. It contains all of \system's code for deploying core services of the system on an arbitrary infrastructure, supported by existing connectors. This repository also contains technical documentation of the system and examples of use cases. 

\smartparagraph{\system's library}. The library of tasks and pipelines implementations is available here: \url{https://github.com/g4allthewaydown/paper-181-library}. This repository contains all tasks, mentioned in this paper, together with other tasks, contributed by the community. We encourage users of the system to freely propose requests to include their tasks and pipeline implementations for public usage in the community.

\smartparagraph{Paper's supplemental materials.} The paper's supplemental materials (such as experiments' code, collected datasets, and required Dockerfiles) are available in this repository: \url{https://github.com/g4allthewaydown/paper-181-supplemental}. While supporting the work described in this paper, this repository will not be used for further system development.
\end{sloppypar}

\end{document}